\documentclass[aip, jcp, twocolumn, floatfix, superscriptaddress, reprint, 10pt]{revtex4-1}  
\newcommand{\vv}[1]{\bm{#1}}

\newcommand{\br}{\mathbf{r}}

\newcommand{\by}{\mathbf{y}}
\newcommand{\bw}{\boldsymbol{w}}

\newcommand{\CX}{\mathcal{X}}

\newcommand{\loss}{\mathcal{L}}

\newcommand{\theSI}{Supplementary Information}
\usepackage[english]{babel}
\usepackage{bm}
\usepackage{graphicx}
\usepackage[usenames]{color}
\usepackage{amsmath}
\usepackage{amssymb}
\usepackage[version=4]{mhchem}
\usepackage[inline]{enumitem}
\usepackage{xfrac}
\usepackage{tabularx}
\usepackage[binary-units=true]{siunitx}
\usepackage{pdfpages}
\makeatletter
\AtBeginDocument{\let\LS@rot\@undefined}
\makeatother

\DeclareSIUnit{\atomicunits}{a.u.}
\DeclareSIUnit{\debye}{D}

\definecolor{mygray}{gray}{0.5}

\usepackage{xspace}

\newcommand{\DefineAuthor}[2]{%
  \expandafter\newcommand\csname #1note\endcsname[1]{%
    \textbf{\textcolor{#2}{\textbf{#1:} ##1}}}%
  \expandafter\newcommand\csname #1\endcsname[1]{%
    \textbf{\textcolor{#2}{##1}}}%
  \expandafter\newcommand\csname #1cancel\endcsname[1]{%
    \textbf{\textcolor{#2}{\sout{##1}}}}%
  \expandafter\newcommand\csname #1change\endcsname[2]{%
    \textbf{\textcolor{#2}{\sout{##1} ##2}}}%
  \newenvironment{#1text}{\color{#2}}{\color{black}}
}
\definecolor{dartmouthgreen}{rgb}{0.05, 0.5, 0.06}
\DefineAuthor{RAD}{red}
\DefineAuthor{YY}{blue}
\DefineAuthor{GEN}{dartmouthgreen}

\begin{document}

\title{Predicting molecular dipole moments by combining atomic partial charges and atomic dipoles}

\author{Max Veit}
\affiliation{Laboratory of Computational Science and Modeling, IMX, \'Ecole Polytechnique F\'ed\'erale de Lausanne, 1015 Lausanne, Switzerland}

\author{David M. Wilkins}
\altaffiliation{Current address: Atomistic Simulation Centre, School of Mathematics and Physics, Queen’s University Belfast, Belfast BT7 1NN, Northern Ireland, United Kingdom}
\affiliation{Laboratory of Computational Science and Modeling, IMX, \'Ecole Polytechnique F\'ed\'erale de Lausanne, 1015 Lausanne, Switzerland}

\author{Yang Yang}
\affiliation{Department of Chemistry and Chemical Biology, Cornell University, Ithaca, NY 14853, USA}

\author{Robert A. DiStasio Jr.}
\email{distasio@cornell.edu}
\affiliation{Department of Chemistry and Chemical Biology, Cornell University, Ithaca, NY 14853, USA}

\author{Michele Ceriotti}
\email{michele.ceriotti@epfl.ch}
\affiliation{Laboratory of Computational Science and Modeling, IMX, \'Ecole Polytechnique F\'ed\'erale de Lausanne, 1015 Lausanne, Switzerland}

\begin{abstract}
The molecular dipole moment ($\boldsymbol{\mu}$) is a central quantity in chemistry.  It
is essential in predicting infrared and sum-frequency generation spectra, as well as induction and long-range electrostatic interactions.  Furthermore, it can be extracted
directly---via the ground state electron density---from high-level quantum mechanical
calculations, making it an ideal target for machine learning (ML).
In this work, we choose to
represent this quantity with a physically inspired ML model that
captures two distinct physical effects: local atomic polarization is captured within the symmetry-adapted
Gaussian process regression (SA-GPR) framework, which assigns a (vector) dipole moment
to each atom, while movement of charge across the entire molecule is captured
by assigning a partial (scalar) charge to each atom.  
The resulting ``MuML'' models are fitted together to reproduce molecular $\boldsymbol{\mu}$ computed using high-level coupled-cluster theory (CCSD) and density functional theory (DFT) on the QM7b dataset, achieving more accurate results due to the physics-based combination of these complementary terms.
The combined model shows excellent transferability when applied to a showcase dataset of larger and more complex molecules, approaching the accuracy of DFT at a small fraction of the computational cost. We also demonstrate that the uncertainty in the predictions can be estimated reliably using a calibrated committee model. The ultimate performance of the models---and the optimal weighting of their combination---depend, however, on the details of the system at hand, with the scalar model being clearly superior when describing large molecules whose dipole is almost entirely generated by charge separation.
These observations point to the importance of simultaneously accounting for the local and non-local effects that contribute to $\boldsymbol{\mu}$; further, they define a challenging task to benchmark future models, particularly those aimed at the description of condensed phases.

\end{abstract}

\maketitle

\section{Introduction}
\label{sec:intro}

The dipole moment $\boldsymbol{\mu}$ of a molecule quantifies the molecule's first-order response to an applied electric field.  It is a key ingredient in the calculation of infrared (IR)\cite{McQuarrie}  and sum-frequency generation (SFG)\cite{Morita2000,rokeNonlinearLight2012} spectra, as well as the understanding of intermolecular interactions.\cite{stone-toif} 
Despite its importance, the dipole moment presents a challenge for calculation, often depending significantly on the level of theory and the basis set used.\cite{DeDroft2000,Hickey2014,haitHowAccurate2018} %
Furthermore, while the molecular dipole moment gives information about the distribution of charge in the molecule, it is determined by the interplay of several physical effects, such as long-range charge transfer and local polarization, which cannot be disentangled based on knowledge of $\boldsymbol{\mu}$ alone.
A number of methods for unravelling these different contributions exist, and are generally based on partitioning the electron density into localized atomic charges and dipoles (accounting for charge transfer and polarization). While these methods are attractive for understanding the underlying physics responsible for $\boldsymbol{\mu}$, they are usually poorly transferable between different molecules or classes of molecules (see Section~\ref{sec:theory}). %

In this work, we design a new framework for the prediction of gas-phase molecular dipole moments that unifies the atomic charge--atomic dipole description rooted in physics with the conformational and chemical sensitivity afforded by kernel-based machine learning (ML).
We begin in Section~\ref{sec:theory} with an overview of existing methods to describe and predict molecular dipoles. 
In Section~\ref{sec:methods}, we formulate the different models we propose to learn and predict polarization:  we use a general symmetry-adapted framework to give environment-centered dipole predictions,~\cite{Grisafi2018} along with a partial-charge model in the vein of existing neural-network models,~\cite{gastegger_machine_2017,Unke2019} to combine good chemical transferability with general conformational dependence.  In Section~\ref{sec:results}, we discuss the training of three models--only partial charges, only environment-centered dipoles, and a combination of the two--which we collectively refer to as MuML.  The models are fitted to reference calculations from high-end linear-response coupled-cluster calculations with single and double excitations (LR-CCSD), and yield $\boldsymbol{\mu}$ with an accuracy that is comparable to that of hybrid density functional theory (DFT).   Next, a showcase set of larger and more complex molecules is used to test these models rigorously.   Finally, we make a critical comparison of the performance of the different MuML models, which reveals the interplay of the different terms that contribute to molecular polarization.

\section{Theory} %
\label{sec:theory}
The molecular dipole moment is defined as the first moment of the total electric charge density,
\begin{equation}
    \boldsymbol{\mu} = - \int\br\,\rho_e(\br)\,{\rm d}^3\br + \sum_{i} \br_i Z_i,
\label{eq:dipole-integral}
\end{equation}
where $\rho_e(\br)$ is the electronic charge density, $\br_i$ the position of the $i^{\rm th}$ nucleus, and $Z_i$ is its charge. 
Usually we are concerned with the permanent dipole moment--that is, the first moment of the total charge density in the molecule's ground-state. However, this expression remains valid for non-equilibrium geometries as well as excited states.

This expression can be simplified by making the approximation that $\rho_e(\br)$ is concentrated at individual atomic sites; that is, each atom $i$ has an associated partial charge $q_{i}$ resulting from the difference between $Z_i$ and the partitioned electron density.  The approximated \textit{total} charge density is thus $\rho(\br) = \sum_{i} q_{i} \delta(\br - \br_{i})$, and we can write
\begin{equation}
\boldsymbol{\mu} = \sum_{i} \br_{i} q_{i},
\label{eq:dipole-charges}
\end{equation}
which is uniquely defined with respect to the origin of the molecular coordinate system if the total molecular charge is zero.
Charged molecules can be accommodated by setting the origin of the molecule to its centroid, such that $\sum_{i} \br_{i} = 0$; this makes the dipole moment invariant to a collective shift of the $q_i$.

The problem then becomes the determination of the $\{ q_i \}$ that best reproduces $\boldsymbol{\mu}$---often in addition to other physicochemical metrics, such as reproducing the molecular electrostatic potential (ESP) or characterizing chemical bonding.  There are many existing methods to determine these charges, varying with the objectives of the model.  Many methods are based directly on the ground-state charge density (or even the wavefunction), such as Mulliken~\cite{mulliken-charges} and L\"{o}wdin~\cite{lowdinNonOrthogonality1950} %
population analyses, 
Hirshfeld decomposition~\cite{hirshfeld-charges} (and its iterative extension~\cite{Bultinck2007}), atoms-in-molecules (AIM, also known as quantum chemical topology--QCT),~\cite{bader-qtaim} and iterative stockholder atoms (ISA).\cite{Lillestolen2008}

Another major class of atomic charge assignment methods, known collectively as ESP fitting methods, focuses directly on reproducing the molecular ESP rather than simply decomposing its charge density. One can immediately see the relevance of such methods to Eq~\eqref{eq:dipole-charges}, as the far-field limit of the electrostatic potential is dominated by the dipolar term.
ESP fitting methods were developed by \citet{momany_determination_1978}, \citet{coxRepresentationMolecular1981}, \citet{Singh1984},and \citet{breneman_determining_1990}; each of these methods finds the charges through a least-squares fit in order to reproduce the ESP at a grid of sites fairly close to the molecule but well outside the van der Waals radius.  Notably, \citet{momany_determination_1978} also fits the total molecular dipole moment in order to satisfy Eq.~\eqref{eq:dipole-charges}. Many subsequent methods incorporate similar information into a fit that makes a compromise between chemical information (the charge density) and far-field electrostatics, such as the DDEC\cite{Manz2012} and Hirshfeld-E\cite{Verstraelen2013} methods.
However, such a compromise becomes a disadvantage when one is only interested in reproducing the molecular dipole moment.

Although the methods above are all motivated by physical and chemical principles, different methods can yield quite different results for the partial charges;\cite{wibergComparisonAtomic1993,stone-toif}
even worse, the results of certain methods may be very sensitive to the details of the underlying electronic structure calculation, such as the basis set used.\cite{coxRepresentationMolecular1981,Singh1984,Stone2005,conwayUseComplete2020}

Furthermore, collapsing the total charge density to a set of points is often too severe an approximation to obtain an accurate description of the ESP.\cite{wibergComparisonAtomic1993,stone-toif}  One can therefore augment the expression in Eq~\eqref{eq:dipole-charges} to include information based on the atom-localized anisotropy or, informally, local polarization of the charge distribution, by adding dipoles (or higher multipole moments) onto the atomic sites.  This is the central idea behind the distributed multipole analysis (DMA) approach,\cite{stone1981} which gives for the total dipole
\begin{equation}
\boldsymbol{\mu} = \sum_{j\in \mathcal{C}} \left( \br_{j} q_{j} + \boldsymbol{\mu}_{j} \right),
\label{eq:dma-mu}
\end{equation}
where $\mathcal{C}$ is a list of centers (or points in real space) that includes both atoms and interatomic positions, $q_{j}$ is the partial charge associated with the $j^{\rm th}$ center, and $\boldsymbol{\mu}_{j}$ is the associated partial dipole. We note in passing that higher multipole moments do not contribute to  $\boldsymbol{\mu}$, and are therefore excluded from Eq~\eqref{eq:dma-mu}.

Several other methods use this idea of representing the molecular ESP with both charges and higher multipole moments assigned to atomic sites, like the FOHI-D model\cite{Geldof2014} and the fullerene polarization model of \citet{mayer_formulation_2007}, the latter recently modified and incorporated into a QM/MM context (where accurate reproduction of the far-field ESP is essential) as the FqF$\mathrm{\mu}$ model.\cite{Giovannini2019}  The authors of FOHI-D in particular separate \emph{intrinsic} atomic polarization, which can be calculated directly for isolated atoms in the same iterative spirit as the classic iterative Hirshfeld method, from atomic \emph{charge transfer}, which is described using the point-charge model.  However, they note that the agreement of their model with the ESP is generally worse when dipoles are included, although this could have been due to their choice of grid points much closer to the molecule than is usually used for ESP-fitting methods.  \citet{mayer_formulation_2007}, on the other hand, discusses the physical idea from the opposite perspective, that of adding atomic charges, derived from a procedure similar to the electronegativity equalization (EEQ) known in chemistry, to an atomic-dipole model in order to describe non-local polarization.  The polarization of carbon nanostructures (nanotubes and fullerenes) is much better described by adding atomic charges to the description, as they can describe the large-scale flow of charge across the conjugated $\pi$-systems typical of these nanostructures.\cite{mayer_formulation_2007}

A key limitation of most of these methods is their inability to describe the dependence of electrostatic quantities across conformational and chemical space without performing additional \emph{ab initio} calculations or fitting empirical parameters,
which severely limits their ability to model experimental spectra and make transferable predictions for new molecules.
A natural way to incorporate the required conformational and chemical sensitivity is to draw on the large body of work over the last two decades that uses ML to predict molecular properties~\cite{darley_beyond_2008,handley_dynamically_2009,Montavon2013,Faber2018,Unke2019,Bereau2018,schuett_schnet_2018,Paruzzo2018,wilkins2019} or molecular and intermolecular potential energy surfaces.~\cite{behler2007,gap-prl,Braams2009,bartok2013,Bartok2017,Morawietz2016,Veit2019}

Many existing methods are explicitly targeted to reproduce $\boldsymbol{\mu}$, or produce it as a side effect.  The earliest of these is the neural network method of \citet{darley_beyond_2008} (see also Ref. \citenum{handley_dynamically_2009}), where a neural network is fitted to reproduce the multipole moments of a molecule or fragment computed via QCT (a.k.a. atoms-in-molecules theory)~\cite{bader-qtaim}.  The two main drawbacks of this strategy, which are common to many of the other methods discussed here, are the following: (1) the need to define a local reference frame, which limits the method's transferability to other chemical compounds, and (2) the need to fit to a precomputed set of atomic charges and multipoles, the choice of which is ultimately arbitrary.  The QCT charges and multipole moments, in particular, are known to be poorly convergent due to the irregular shapes of the partitioned atomic volumes.~\cite{stone-toif}

Techniques for fitting local electrostatic properties have evolved considerably since then, but most of the proposed methods retain these two key drawbacks.  For example, the IPML model of \citet{Bereau2018} predicts intermolecular interaction energies accurately by systematically treating several different physical energy contributions.  The dipole moments themselves, on the other hand, are not as well predicted, given that their accurate reproduction is not the primary goal of the model.  Part of the error may have come from using environment-local axis systems to predict the higher-order multipole moments, which is a less general and robust approach than the symmetry-adapted regression introduced in \citet{Grisafi2018}. Furthermore, the model retains the same drawback of being fitted to a specific partitioning scheme---in this case, the minimal-basis iterative stockholder method~\cite{verstraelen_minimal_2016}, which was chosen for its accuracy in modelling electrostatic interactions and not for reproducing $\boldsymbol{\mu}$.

The neural network model of \citet{gastegger_machine_2017}, on the other hand, does explicitly target $\boldsymbol{\mu}$.  It predicts the set of environment-dependent partial charges that best reproduces the total dipole moment, thereby bypassing the need to choose an arbitrary charge partitioning scheme, and uses the conformational sensitivity gained through the neural network to accurately predict infrared spectra.  The PhysNet model of \citet{Unke2019} uses the same idea and additionally uses a new representation to span a large swath of chemical space, as does the HIP-NN model of \citet{sifainDiscoveringTransferable2018}, which also incorporates enough conformational dependence to be able to predict infrared spectra.  All three of these models only predict scalar atomic properties, neglecting contributions from  atomic polarization, which we will see are important to achieving the best accuracy and transferability.%

There are several approaches to fitting properties, such as $\boldsymbol{\mu}$, that transform as tensors--in particular, approaches that are covariant (rather than invariant) to rotations.  The local-axis approach used in \citet{Bereau2018} has already been mentioned; another approach is the covariant kernels introduced in \citet{glielmoAccurateInteratomic2017} and developed into a general symmetry-adapted regression method for any tensor order in \citet{Grisafi2018}.  (Covariant kernels were tried in \citet{Bereau2018} but found to perform about as well as the local-axis method, although this is not the same formulation as that used in \citet{Grisafi2018}.)  In this regard, the symmetry-adapted regression method was successfully tested on dipole moments of small molecules and clusters, as well as accurately predicting higher-order tensors such as the polarizability~\cite{wilkins2019}.

Finally, \citet{Christensen2019} have developed a formalism (OQML) for incorporating electric field gradients into a ML fit.  They use a system of arbitrary, though usually realistic, partial charges in order to define an implicit local reference frame for each atomic environment, which can then be used to fit local dipole moments.  While their formulation is quite different from the method developed below, we believe it is fundamentally similar to assigning an environment-dependent partial dipole to each atom, as described in Section~\ref{sec:partial-dipole-model}.

\section{Methods}
\label{sec:methods}

\subsection{Partial-Charge Model}
\label{sec:partial-charge-model}

We begin by building a ML model that incorporates local environment sensitivity into the simple partial-charge model of Eq.~\eqref{eq:dipole-charges} using Gaussian process regression (GPR)\cite{rasmussenGaussianProcesses2006}. To do this, we exploit the fact that GPR uses a linear fit in kernel space, and can therefore be used to fit the result of any linear operator applied to atomic quantities~\cite{ceri+18book}. The vector of weights $\bw$ is required that minimizes the regularized loss function,
\begin{equation}
    \loss^2 = \| \mathbf{L}\mathbf{K}_{PM}\bw - \by \|^2_{\Lambda^{-1}} + \|\bw\|^2_{\mathbf{K}_{MM}}
\end{equation}
where $\Lambda$ is a diagonal matrix whose entries $\sigma^{2}_{\mu}$---a quantity known as the ``dipole regularization'', usually kept the same for all molecules---are chosen to optimize the error of the fit along with its transferability to new molecular databases, $\mathbf{L}$ a linear operator, and
\begin{equation}\label{eq:pcharge_weights}
    \bw = (\mathbf{K}_{MM} + (\mathbf{L}\mathbf{K}_{PM})^T \Lambda^{-1} \mathbf{L}\mathbf{K}_{PM})^{-1} \mathbf{L}\mathbf{K}_{PM} \Lambda^{-1} \by.
\end{equation}
The fit uses an ``active set'' of $M$ basis functions (which in practice is a
small fraction of the total number $P$ of atoms in the database). Following the same notation introduced in Ref.~\citenum{ceri+18book}, we use $M$ and $P$ to indicate both the sets and the number of entries. The kernel matrices $K_{MM}$ and $K_{PM}$ contain the kernel evaluated between all sparse points ($M$) and themselves, as well as with all atoms in the training database ($P$).
In
principle, any sufficiently representative set of configurations could be used
to form the active set of basis functions, but in practice they are almost
always chosen from the environments present in the molecules in the training
set using an algorithm such as farthest point sampling (FPS) or
a CUR decomposition.\cite{imbalzanoAutomaticSelection2018}.
The entries of the kernel matrix are $(K_{IJ})_{ij} = k(\CX^{(i)},\CX^{(j)})$,
where the SOAP kernel~\cite{bartok2013} is used as the similarity function
$k(\cdot,\cdot)$ between two atomic environments $\mathcal{X}^{(i)}$ and $\mathcal{X}^{(j)}$.

To build up a model for $\boldsymbol{\mu}$, we predict partial charges
$q(\mathcal{A}_{i})$ for atom $i$ in molecule $\mathcal{A}$ 
\begin{equation}
q(\mathcal{A}_{i}) = \sum_{j \in M} w_{j} k(\CX^{(i)},\CX^{(j)})
\label{eq:partial-charge-prediction}
\end{equation}
(where the sum runs over all basis points, i.e. environments in the active set), such that
\begin{equation}
\vv{\mu}(\mathcal{A}) = \sum_{i\in\mathcal{A}} \br_{i} q(\mathcal{A}_{i}) .
\label{eq:partial-charge-atomsum}
\end{equation}
We can then define the transformed kernel matrix between dipoles and basis points as
\begin{equation}\label{eq:linear_transform_matrix_pcharge}
(\mathbf{L}_\mu\mathbf{K}_{PM})_{\mathcal{A},j} = \sum_{i\in\mathcal{A}} \br_{i} k(\CX^{(i)},\CX^{(j)}),
\end{equation}
allowing us to use Eq.~\eqref{eq:pcharge_weights} to determine the weights.
The expression in Eq.~\eqref{eq:linear_transform_matrix_pcharge} represents a block of 3 rows from the $3N\times M$ matrix $\mathbf{L}_\mu\mathbf{K}_{PM}$,  where $N$ is the number of molecules in the training set, one row for each Cartesian component of the dipole moment of molecule $\mathcal{A}$. The columns index  $j$ runs over the  $M$  environments in the active set.  %
The $\br_i$ are defined with respect to the coordinate system in which the dipole is given, with the origin set to the centroid of the respective molecule so that the prediction is insensitive to a shift in the total charge.  The target data $\by$ are then defined as a concatenation of the Cartesian components of the training-set dipole moments.

The insensitivity of the model to the total molecular charge is advantageous because the model's total charges, $Q_\mathcal{A} = \sum_{i \in \mathcal{A}} q_i$, need not be constrained to reproduce exactly the total molecular charge.  
As noted in \citet{Unke2019}, applying this constraint to the training set would not guarantee that the model gives the correct charges for prediction on a new molecule.
Furthermore, we found that including exact total-charge constraints into the
fit via Lagrange multipliers severely reduced the quality of the fit---in most
cases simply giving all partial charges as zero---because the procedure used
to select the sparse active set of $M$ environments also discarded the basis
functions necessary to satisfy this constraint whilst also satisfactorily
reproducing $\boldsymbol{\mu}$.

However, it is usually beneficial to include some sort of restraint (even if not an exact \emph{constraint}) %
on total charge, as a model insensitive to this quantity can predict unreasonably large total charges, ultimately compromising its transferability to other datasets.  We therefore include the total charge as extra information to the fit by appending to $\by$ the list of total charges of the molecules in the training set, and appending to the transformed kernel matrix $\mathbf{L}_\mu\mathbf{K}_{PM}$ the extra $N$ rows representing the sums of the model's partial charges:
\begin{equation}
    (\mathbf{L}_Q\mathbf{K}_{PM})_{\mathcal{A},j} = \sum_{i \in \mathcal{A}} k(\CX^{(i)}, \CX^{(j)}) ,
    \label{eq:charge-restraint-matrix}
\end{equation}
and extending the diagonal regularization matrix $\Lambda$ with an extra $N$ entries $\sigma^2_Q$---the charge regularizer---in order to be able to regularize the two target quantities separately. %

\subsection{Partial Dipole Model}
\label{sec:partial-dipole-model}

An alternative method for predicting $\boldsymbol{\mu}$ is to build up the prediction as a sum of atom-centered dipole moment predictions using symmetry-adapted Gaussian process regression (SA-GPR)\cite{Grisafi2018}, a modification of standard GPR that allows tensor properties to be learned.
A SA-GPR prediction of the dipole moment $\boldsymbol{\mu}$ of a test molecule $\mathcal{A}$ is given by:
\begin{equation}
\boldsymbol{\mu}(\mathcal{A}) = \sum_{j\in M}\sum_{i\in\mathcal{A}}\mathbf{k}^{V}(\mathcal{X}^{(i)},\mathcal{X}^{(j)}) \bw_j,
\label{eq:dipole-lambda-soap}
\end{equation}
where $\mathbf{k}^V(\mathcal{X}_{i},\mathcal{X}_{j})$ is an element of an extended kernel matrix, being the tensor (concretely a $3\times 3$ matrix) whose components $k_{\alpha\beta}^{V}(\mathcal{X}_{i},\mathcal{X}_{j})$ give the coupling between the   Cartesian component $\mu^{(i)}_\alpha$ associated with environment $\mathcal{X}_{i}$ and the $\mu^{(j)}_\beta$ component associated with $\mathcal{X}_{j}$.  Each environment $j$ in the active set now requires a set of three weights (represented by the vector $\bw_j$) to represent the three independent components of the vector quantity assigned to each atom.

Since the dipole moment is a vector quantity that 
is related by a linear transformation to the spherical harmonics with $L=1$,
the vector kernel $\mathbf{k}^{V}(\mathcal{X},\mathcal{X}')$ can be obtained directly from the $\lambda=1$-order $\lambda$-SOAP kernel of SA-GPR, $\boldsymbol{k}^{\lambda=1}(\mathcal{X},\mathcal{X}')$ by the transformation,
\begin{equation}
\boldsymbol{k}^{V}(\mathcal{X},\mathcal{X}') = \boldsymbol{M}^{\dagger} \boldsymbol{k}^{\lambda=1}(\mathcal{X},\mathcal{X}') \boldsymbol{M},
\end{equation}
where $\mathbf{M}$ transforms from the Cartesian basis to the basis of $\lambda=1$ spherical tensors (see e.g. Ref.~\citenum{stone-toif} for an explicit formula).

As has been shown by recent work,\cite{Grisafi2018,wilkins2019,raimbaultUsingGaussian2019} SA-GPR performs very well for response properties of different orders in a wide variety of systems.  Further, one can see from the atom-centered formulation of Eq.~\eqref{eq:dipole-lambda-soap} that the atom-centered dipoles, analogous to the atomic partial charges of Eq.~\eqref{eq:partial-charge-prediction}, can easily be extracted:
\begin{equation}
\boldsymbol{\mu}(\mathcal{A}_{i}) = \sum_{j\in M}\mathbf{k}^{V}(\mathcal{X}^{(i)},\mathcal{X}^{(j)}) \bw_j.
\label{eq:vector-peratom-dipoles}
\end{equation}
Although an SA-GPR prediction of $\boldsymbol{\mu}$ does not require charge
constraints, it is computationally more expensive than a partial-charge model,
requiring the inversion of a square matrix with three times the number of rows
($3M$ rows, where $M$ is the number of basis functions in the active set).

\begin{table*}[bth]
    \centering
    \sisetup{table-number-alignment=center}
    \begin{tabular}{l
        S[table-format=1.6]
        S[table-format=1.6]
        S[table-format=1.6]
        S[table-format=1.6]
        S[table-format=1.6]}
        Model & $\theta_a / \si{\angstrom}$ & $r_{0} / \si{\angstrom}$ & $m$ & $\sigma_\mu / {}$\num{e-3} & $\sigma_Q / {}$\SI{e-3}{\per\angstrom} \\ \hline
        Scalar (CCSD) & 0.375 &  2.32 & 4.41 & 4.38 & 35.5 \\
        Scalar (B3LYP) & {--} & {--} & {--} & 4.41 & 78.8 \\
        Vector (CCSD) & 0.256 & 2.75 & 3.34 & 1.47 & {--} \\
        Vector (B3LYP) & {--} & {--} & {--} & 1.15 & {--} 
    \end{tabular}
    \caption{Optimal hyperparameters for the pure scalar and pure vector models, obtained using a Nelder-Mead optimization.  $\theta_a$: Gaussian width for SOAP atom smearing, $r_{0}$ and $m$: radial scaling parameters (see Ref.~\citenum{willattFeatureOptimization2018}), $\sigma_\mu$: dipole regularization (unitless since $\delta = 1$ for the pure fits), $\sigma_Q$: total charge regularization.   Parameters for the combined model are derived as indicated in the text.  All numbers truncated to three significant figures. %
    }
    \label{tab:kopt_results}
\end{table*}

\begin{table}[bth]
    \centering
    \sisetup{table-number-alignment=center}
    \begin{tabular}{l c c c c}
        Model & $n_\text{max}$ & $l_\text{max}$ & $N_F$ & $M$ \\
        \hline
        Scalar & 8 & 6 & 200 & 2000 \\
        Vector & 4 & 2 & 200 & 2000
    \end{tabular}
    \caption{Convergence parameters for the scalar and vector kernels: $n_{max}$ is the number of radial basis functions and $l_{max}$ is the angular momentum band limit for the SOAP kernel, $N_F$ is the number of selected sparse features, and $M$ is the number of selected sparse environments for each model.  Note that the scalar and tensor \emph{power spectrum} components of the vector SOAP kernel use the same parameters.}
    \label{tab:kernel_convergence_params}
\end{table}

\subsection{Combined Model}
\label{sub:combined-model}

We now consider the partial-charge model and partial-dipole model as two separate models for the same system, encoding two different physical effects.  It should then be possible to get a better prediction simply by fitting the \emph{sum} of the two models to the training data.  We call the matrix of Eq.~\eqref{eq:linear_transform_matrix_pcharge} the ``transformed scalar kernel'':
\begin{equation}
    \mathbf{K}^S_{NM} := \mathbf{L}_\mu\mathbf{K}_{PM},
    \label{eq:transformed-scalar-kernel}
\end{equation}
and the analogous ``transformed vector kernel'' $\mathbf{K}^V_{NM'}$ whose rows are the atom-wise summations of the kernel from Eq.~\eqref{eq:dipole-lambda-soap}:
\begin{equation}
    (\mathbf{K}^V_{NM'})_{\mathcal{A},j} = \sum_{i\in \mathcal{A}}\mathbf{k}^V(\mathcal{X}^{(i)}, \mathcal{X}^{(j)}).
\end{equation}

Because the models are of different dimensions and model different physical effects, we assign each a different weight vector: $\bw^S$ for the scalar weights and $\bw^V$ for the vector weights.  Note also that this means we do not need to use the same set of basis functions for the scalar and vector models; they can be chosen independently.  Then, in order to find the best combined sum model, we optimize
\begin{multline}
    \loss^2 = \|\delta_S^2 \mathbf{K}^S_{NM}\bw^S + \delta_V^2 \mathbf{K}^V_{NM'}\bw^V - \by\|_{\Lambda^{-1}}^2 \\
    + \delta_S^2 \|\bw^S\|_{\mathbf{K}^S_{MM}}^2 + \delta_V^2\|\bw^V\|_{\mathbf{K}^V_{M'M'}}^2
    \label{eq:combined-model-loss}
\end{multline}
with respect to both sets of weights $\boldsymbol{w}^S$ and $\boldsymbol{w}^V$ simultaneously.  (The $\mathbf{K}^S_{MM}$ and $\mathbf{K}^V_{M'M'}$ are the matrices of non-transformed kernels of all the basis functions with each other.) The result can be expressed using the inversion of a square matrix with $M + M'$ rows, where $M'$ is the number of vector weights (three times the number of vector basis functions).  Since the number of basis functions is usually kept the same for both scalar and vector models, the matrix to be inverted has $4M$ rows, making the combined model the most expensive of the three models discussed here.  In practice, however, the cost is typically manageable.

Furthermore, the charge restraint %
can be incorporated as discussed in Section~\ref{sec:partial-charge-model}, where the transformed scalar kernel is appended with the matrix from Eq.~\eqref{eq:charge-restraint-matrix} and the transformed vector kernel is appended with the same number of rows of zeros (since the vector model does not contribute to the total molecular charge).
Note also that we have introduced weights $\delta_S$ and $\delta_V$ to modify the overall relative amount that the scalar and vector components contribute to the combined model. The $\delta$-weights effectively allow for different regularizations of the scalar and vector components of the model, which is equivalent to assuming different variances for the dipole components modelled by the scalar and vector models~\cite{bartok2015}.

\section{Results and Discussion}
\label{sec:results}

We optimized and trained the scalar, vector and combined models on the QM7b data set,\cite{Rupp2012QM7} which contains $\num{7211}$ small organic molecules with up to seven heavy/non-hydrogen atoms (specifically C, N, O, S, and Cl) with varying degrees of H saturation. The dipoles were computed using the methods described in \citet{yangQuantumMechanical2019}, namely DFT with the hybrid B3LYP functional~\cite{beckeDensityFunctional1993,kimComparisonDensity1994} and linear-response coupled-cluster theory with single and double excitations (LR-CCSD~\cite{kochCoupledCluster1990}, hereafter just `CCSD'). In both cases, the doubly augmented double-$\zeta$ d-aug-cc-pVDZ basis set\cite{woonGaussianBasis1994} (hereafter referred to as `daDZ') was employed during all calculations. 
We then demonstrate the transferability of this model on the QM9\cite{Ramakrishnan2014} data set, comparing with state-of-the-art results from Ref.~\citenum{Christensen2019}, and on a ``MuML showcase'' data set of larger molecules. %
Finally, we push the models to their limits by studying different polymers composed of or derived from the glycine amino acid. 

\subsection{Model Optimization}
\label{sub:model-optimization}
We first optimize the models for space and computational requirements by subsampling the SOAP feature matrices (which are multiplied and raised to an integer entry-wise power to obtain the SOAP kernel) using the FPS selection algorithm described in \citet{imbalzanoAutomaticSelection2018}.  Descriptors are first subsampled in the feature space dimension, allowing for fewer SOAP components ($N_F$) to be used in calculating the kernel, then in the environment space dimension, allowing for fewer representative environments ($M$) to be used when performing the fit.  The convergence of the final fitting error with respect to these parameters, as well as other kernel convergence parameters such as the number of radial channels ($n_\text{max}$) and the maximum angular momentum ($l_\text{max}$) of the expansion, is shown in the \theSI{}.

We chose the model's overall distance-based cutoff as \SI{5}{\angstrom}, to encompass all atom pairs in the QM7b dataset.  The actual radial dependence of the kernels, however, is optimized using the radial-scaling function from \citet{willattFeatureOptimization2018}.  Together with the SOAP atom width and the regularization parameters, this leaves us with several continuous hyperparameters whose optimal values need to be determined.  In a Bayesian approach, these would be considered priors; they would ideally be integrated over using a previously-known prior distribution.  
Here, however, we do not have much prior knowledge about the distribution of these parameters---in contrast to the study of potential energy surfaces, where good values can be guessed quite accurately based on prior experience and physical knowledge.\cite{Bartok2018}  Instead, we use optimization to find the best values of these parameters for our problem, along with cross-validation (CV) to guard against the problem of overfitting (which is otherwise introduced by hyperparameter optimization techniques).

First, the hyperparameters for the scalar and vector models are each independently optimized on a randomized four-fold CV split of $\num{5400}$ randomly-selected molecules of the QM7b test set\cite{Rupp2012QM7}.  The results of this optimization can be found in Table~\ref{tab:kopt_results}.  
The combined model is then obtained as follows: since there are only three free parameters between the overall scalar weight $\delta_S$, the overall vector weight $\delta_V$, the dipole regularization $\sigma_\mu$, and the total charge regularization $\sigma_Q$, we set the dipole regularization to 1 and scale the rest of the parameters  accordingly: If $\sigma_\mu^S$ is the optimal dipole regularizer and $\sigma_Q^S$ the optimal charge regularizer for the scalar model, and if $\sigma_\mu^V$ is the optimal dipole regularizer for the vector model, then we take $\delta_S \mapsto \sfrac{1}{\sigma_\mu^S}$, $\sigma_Q \mapsto \sfrac{\sigma_Q^S}{\sigma_\mu^S}$, and $\delta_V \mapsto \sfrac{1}{\sigma_\mu^V}$.
Further details of the optimization procedure are discussed in the \theSI{}.

Finally, once the model's hyperparameters are converged and optimized, model training and testing are quite fast.  For example, computing scalar and vector training and testing kernels for the set of $N=\num{20000}$ molecules of QM9 used in Section \ref{sub:qm9-test} (with a test set of $T=\num{1000}$ molecules; $M=M'=\num{2000}$) required just over 1 hour and \SI{95}{\gibi\byte} of memory on a modern 24-core machine, with almost all of the time and memory used to compute the training kernels; the test kernels required less than 2 minutes and \SI{3}{\gibi\byte}.  Once the kernels were computed, fitting the combined (most expensive) model required only 2 minutes and \SI{20}{\gibi\byte} of memory, and computing test-set predictions was almost negligible in comparison, taking 2 seconds and \SI{1}{\gibi\byte} of memory.  This means that the regularizers can be optimized quite cheaply once optimal kernels have been computed.

\subsection{Error measures}

Throughout this work, we use two different error measures. The ``per-atom'' RMSE (root-mean-squared error)
\begin{equation}
    \text{RMSE} = \sqrt{\frac{1}{N_\text{test}}\sum_{j \in \text{test}} \left\|\frac{\boldsymbol\mu_\text{predicted}^{(j)} - \boldsymbol\mu_\text{actual}^{(j)}}{N_j}\right\|_2^2},
    \label{eq:per-atom-rmse}
\end{equation}
reports on both the magnitude and the orientation of the predicted dipoles.
The residuals are normalized by the number of atoms $N_j$ in the respective molecule before taking the RMSE. This scaling posits a generally linear trend of the dipole moment norm as a function of the number of atoms.  Such a trend would be expected from an additive model where each atom contributes a certain, locally-dependent amount.  This is the case with the vector model, but not with the scalar model, where the contribution additionally depends on its distance from the molecular origin, making the scaling depend on the molecular geometry. 
Therefore, to provide an alternate assessment of the error of the total dipole, and to facilitate comparison with other studies, we additionally plot the MAE (mean absolute error) of the norm of the \emph{total} dipole moment:
\begin{equation}
\text{MAE} = \frac{1}{N_\text{test}} \sum_{j \in \text{test}} \Big| \|\boldsymbol\mu_\text{predicted}^{(j)}\|_2 - \|\boldsymbol\mu_\text{actual}^{(j)}\|_2 \Big|.
    \label{eq:dipole-norm-mae}
\end{equation}
For the QM7b dataset, these two measures provide similar information, but for transferability testing on other datasets these measures provide complementary information.

\subsection{Uncertainty quantification}\label{sec:uq}

We can estimate the uncertainty in the model predictions using a calibrated committee model, as described in \citet{musilFastAccurate2019}. We train $n_{\text{comm}}$ models $\tilde{\boldsymbol\mu}^{(k)}(\mathcal{A})$, using the same active set but choosing a different random subset of the full training set in each model. The predictions of these models are then rescaled around their mean 
\begin{equation}
\begin{split}
\bar{\boldsymbol{\mu}}(\mathcal{A}) = &
\frac{1}{n_{\text{comm}}} \sum_k \tilde{\boldsymbol\mu}^{(k)}(\mathcal{A}), \\
\boldsymbol\mu^{(k)}(\mathcal{A})=&
\bar{\boldsymbol{\mu}}(\mathcal{A})
+\alpha\left(\tilde{\boldsymbol\mu}^{(k)}(\mathcal{A})-
\bar{\boldsymbol{\mu}}(\mathcal{A})
\right)
\end{split}
\end{equation}
by a calibration factor $\alpha$, that is determined using the ``internal validation procedure'' described in Ref.~\citenum{musilFastAccurate2019}.  The best estimate of the committee model is given by its mean, $\bar{\boldsymbol{\mu}}(\mathcal{A})$, and uncertainty is then computed as the standard deviation of the rescaled predictions. Individual members of the calibrated committee can be used to separately compute derived quantities (e.g., the norm of the dipole moment), which greatly simplifies the propagation of uncertainty (see e.g. Ref.~\citenum{raimbaultUsingGaussian2019}). While the use of a committee model for a sparse Gaussian process model entails virtually no computational overhead when making a new prediction, the training process is somewhat more cumbersome. For this reason, we only use a committee model when making predictions for the showcase dataset in Section~\ref{sub:showcase}. More systematic tests performed on benchmark datasets use a single regression model, without error estimation, which usually also achieves a higher accuracy than the ensemble average (see the \theSI{}) because it is trained on all training points together.

\begin{figure}[hbtp]
    \centering
    \includegraphics[width=\columnwidth]{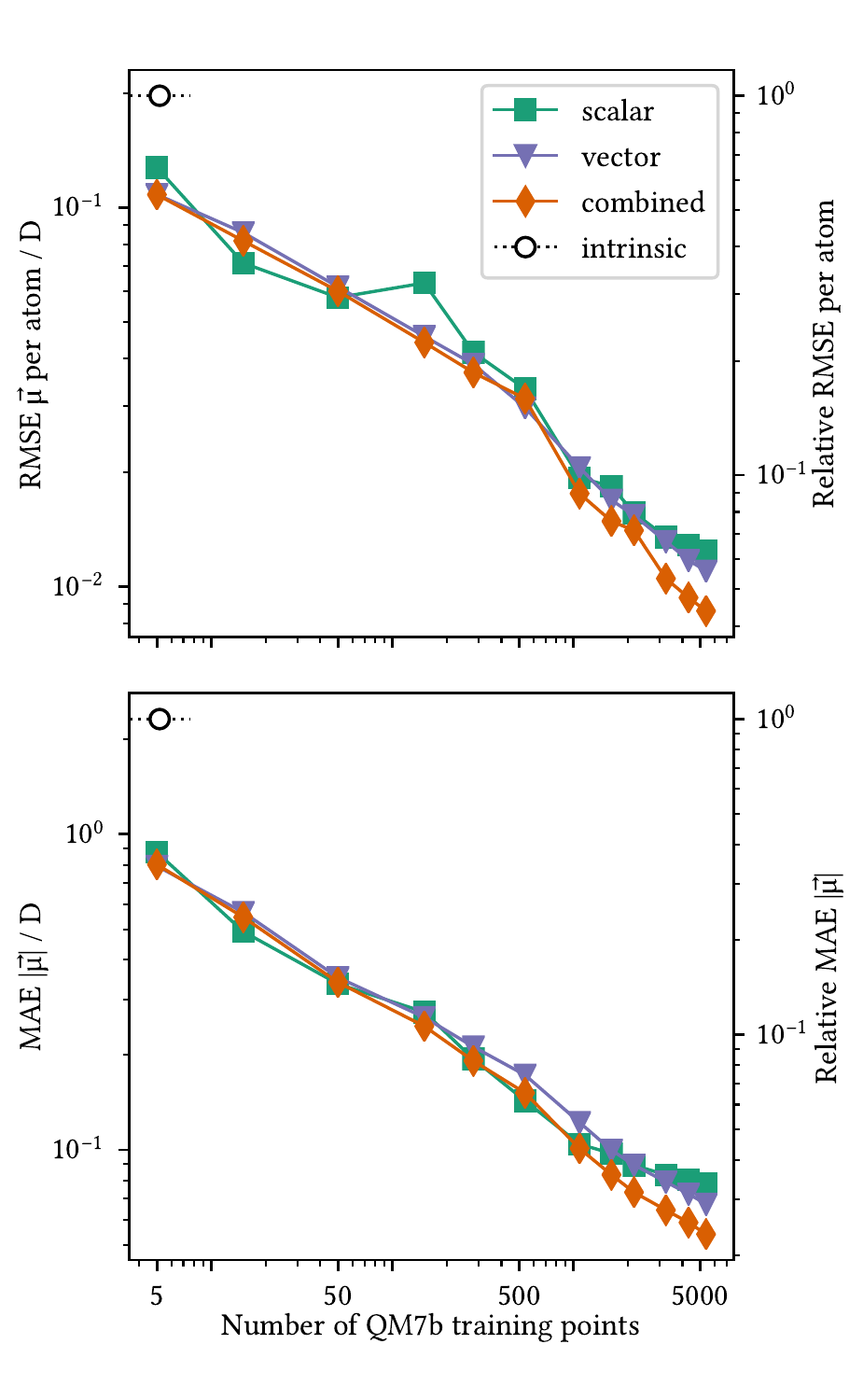}
    \caption{Learning curves of the scalar (green squares), vector (purple triangles), and combined (orange diamonds) models for CCSD/daDZ dipoles computed on 1811 randomly-selected molecules of the QM7b dataset.  The models were trained on subsets of the remaining 5400 molecules.  The top plot has per-atom RMSEs and the bottom plot has per-molecule dipole moment norm MAEs.  The open circle denotes the intrinsic variation of the dataset, i.e., the error of a zero model.}
    \label{fig:lcs_qm7}
\end{figure}

\subsection{Training on QM7b}

Figure~\ref{fig:lcs_qm7} shows the learning curves of the MuML models, with the kernel parameters fixed to the values optimized on 5400 points. Errors are computed on a test set of 1811 randomly-selected molecules from the QM7b dataset~\cite{Rupp2012QM7}.  Note that the pure scalar and pure vector models both achieve similar performance in the limit of a large amount of data, while the combined model clearly outperforms both (by a factor of about 20\%) in the same regime.

This figure reports results for models trained on CCSD/daDZ dipoles. Results for B3LYP/daDZ-trained models are very similar (see SI).
For reference, the discrepancy between B3LYP/daDZ and CCSD/daDZ molecular dipole moments in the QM7b database amounts to an $\mathrm{RMSE} = {}$\SI{0.011}{\debye} per atom, or $\mathrm{MAE} = {}$\SI{0.087}{\debye}. It should be stressed that, contrary to the case of the polarizability~\cite{wilkins2019,yangQuantumMechanical2019}, the performance of DFT is usually quite satisfactory when predicting molecular dipole moments. When trained on 5400 QM7b structures, the combined model delivers better accuracy ($\mathrm{RMSE} = {}$\SI{0.0086}{\debye} per atom, $\mathrm{MAE} = {}$\SI{0.054}{\debye}), at a dramatically reduced computational cost.

\begin{figure}[h!!!]
    \centering
    \includegraphics[width=\columnwidth]{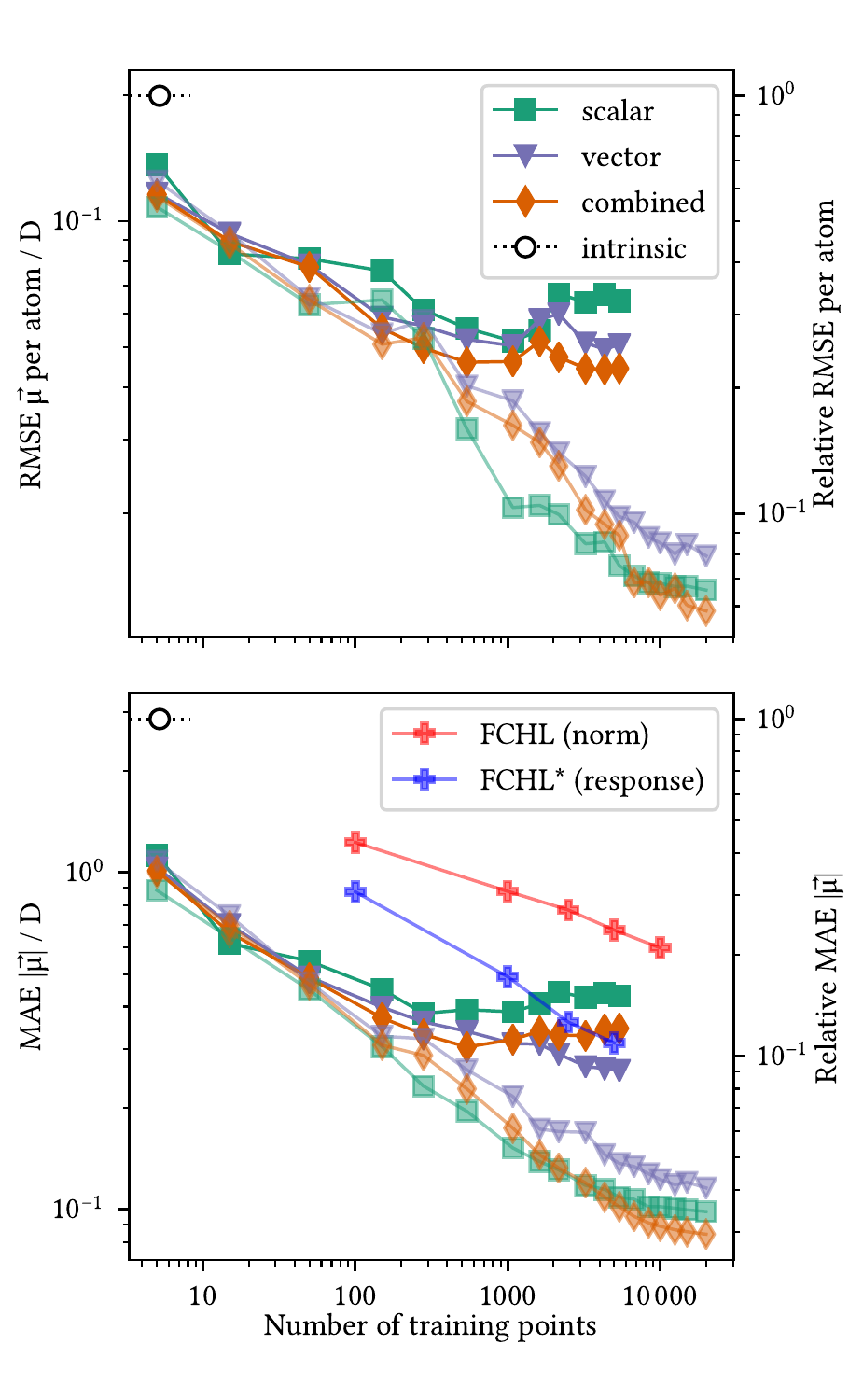}
    \caption{Learning curves on a random sample of 1000 molecules from the QM9 dataset\cite{Ramakrishnan2014}.  Reference dipoles were computed with B3LYP/daDZ; MuML models were re-trained on QM7b dipoles (solid symbols) and QM9 dipoles (semi-transparent symbols) computed at the same level of theory.  Top: per-atom RMSEs.  The QM7b combined fit narrowly outperforms the pure charge and pure dipole models, with significant saturation apparent in all QM7b models.  No such saturation is apparent in the QM9 models.  Bottom: MAE of the error of the dipole moment norms for each molecule.  The FCHL (norm-only) and FCHL* (vector response) curves are reproduced from \citet{Christensen2019}; both models were trained on QM9 dipoles.  Using this error measure, the QM7b pure vector fit has a clear advantage, even outperforming the FCHL* response learning.  The QM9 fits again perform significantly better than the QM7b fits; the QM9 combined fit retains the best performance, reaching an MAE of \SI{0.084}{\debye} at \num{20000} training points.
    }
    \label{fig:lcs_qm9}
\end{figure}

\begin{figure}
    \centering
    \includegraphics[width=\columnwidth]{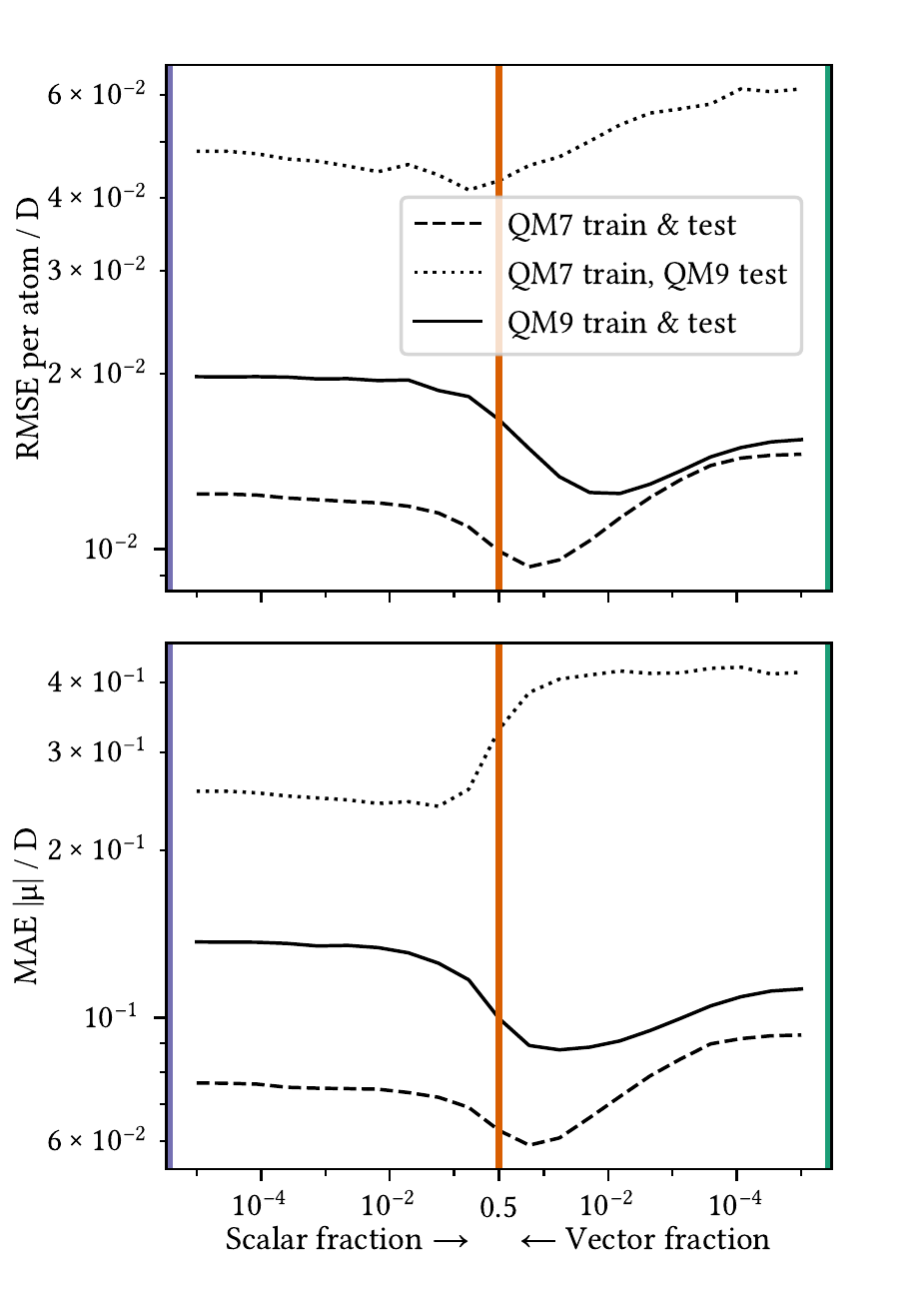}
    \caption{Errors when interpolating between the pure scalar and pure vector B3LYP/daDZ MuML models.  The pure vector model is located at the left (as the scalar fraction $t \rightarrow 0$), the pure scalar model is located at the right (as the vector fraction $1-t \rightarrow 0$), and the combined model is located in the middle at \num{0.5}.  The models are trained on either $\num{5400}$ points of QM7b or $\num{5400}$ points of QM9, and tested on either the remaining $\num{1811}$ molecules of QM7b or the $\num{1000}$ randomly selected molecules of QM9.}
    \label{fig:scalar_vector_interp}
\end{figure}

\subsection{Testing on QM9}
\label{sub:qm9-test}

In order to test the extrapolation capabilities of the MuML models, we selected $\num{1000}$ random samples from the QM9 dataset\cite{Ramakrishnan2014} and computed the dipole moments following the same protocol used for the QM7b dataset~\cite{wilkins2019,yangQuantumMechanical2019}.
Due to the high computational cost of CCSD, we used B3LYP/daDZ as the reference in this case, and the corresponding models trained on QM7b at the B3LYP/daDZ level.
The learning curves of these models are shown in Figure~\ref{fig:lcs_qm9}.  The combined model outperforms the scalar and vector models in terms of the per-atom RMSE measure, but performs worse than the vector model using the norm MAE. The errors are much larger than those seen when testing on QM7b, and the asymptotic behavior of the learning curves indicates saturation and even overfitting. 
In order to determine whether the saturation in model performance is due to limitations in the models, or just insufficient training data, we also computed learning curves for models trained on QM9 dipoles, using a set of \num{20000} additional molecules drawn from the QM9 set and dipoles computed at the B3LYP/daDZ level.  
The scalar and vector regularizers were re-optimized using \num{15000} training points.  The QM9-trained models, in contrast to the QM7b-trained models, do not saturate early.  The QM9-trained combined model reaches an MAE of \SI{0.084}{\debye}; this is more accurate than the QM9-trained scalar model (MAE \SI{0.099}{\debye}), which is in turn more accurate than the QM9-trained vector model (MAE \SI{0.12}{\debye}).  In fact, this is comparable to the performance of the SchNet neural network model~\cite{schuett_schnet_2018}, which reaches an accuracy of \SI{0.033}{\debye} using \num{110000} training molecules.  It is likely that the QM9-trained combined model would reach the same accuracy if the slight saturation in the MAE curve were corrected, e.g., by increasing the SOAP convergence parameters (which were set for 5400 QM7b molecules) and re-optimizing the hyperparameters.
The comparison of the QM7b-trained models to those trained on QM9 clearly shows that the QM7b scalar model especially suffers in the extrapolative regime.  Together with the degrading performance of both RMSE and MAE as the number of training points approaches the full training set size, this indicates that the scalar model has a strong tendency to overfit.  As for the combined model, it seems that its poor performance is a result of its inclusion of too much of the overfitted scalar component.

We therefore investigate the dependence of the model error on the scalar-vector mixing, to see if the combined model can be improved by including less of the scalar component.  The scalar-vector mixing is parametrized here by varying the scalar model's variance $\delta^2_S$ from zero to its pure-scalar equivalent value: $\delta^2_S(t) = t(\sigma_\mu^S)^{-2}$ whilst simultaneously varying the vector model's variance from its pure-vector equivalent value to zero: $\delta^2_V(t) = (1 - t)(\sigma_\mu^V)^{-2}$.  The dipole regularization is kept at one, as the regularization is encoded in the model variances; the total-charge regularization $\sigma_Q$ is likewise kept constant, at its optimal scalar-model value, as it only applies to the scalar model.  This parametrization reproduces the pure scalar and pure vector predictions at each endpoint whilst smoothly transferring the total model's variance from the vector to the scalar model.  The value of $t=0.5$ corresponds to the combined model (modulo a factor of 2 in the regularizer, which is negligible in practice).  Note that varying the scalar and vector weights $\delta_S$ and $\delta_V$ is more than a simple post-processing adjustment; it requires recomputing the model weights $\boldsymbol{w}$ via Eq.~\eqref{eq:combined-model-loss} as well.
We plot such a scalar-vector scan in Figure~\ref{fig:scalar_vector_interp} for the models trained on either QM7b or QM9, and tested on either QM7b (QM7b-trained model only) or QM9 (both models).  Both models were trained on \num{5400} molecules using dipoles computed at the B3LYP/daDZ level.  We see that the optimum for models tested in the interpolative regime---that is, QM7b tested on QM7b and QM9 tested on QM9---does not in fact lie at $t=0.5$, but closer to the pure-scalar model (vector fraction of 0.1 or 0.01, depending on the model and whether one wants to optimize MAE or RMSE). The na\"ive QM7b combined model at $t=0.5$ is still better than either the pure scalar or pure vector models (this is also the case with the QM9 model once we add more training points).  On the other hand, for the QM7b model in its extrapolative regime (i.e. tested on QM9), the situation is the opposite: the optimal model has a \emph{scalar} fraction of around 0.1, and the na\"ive combined model at $t=0.5$ is even worse in MAE than the pure vector model, as we have seen in Figure~\ref{fig:lcs_qm9}.

These observations confirm our suspicions that the scalar model is prone to overfitting, as it achieves very good performance in the interpolative regime, but relatively poor performance in the extrapolative regime.  Models with a higher fraction of the vector contribution, on the other hand, may not achieve the same accuracy in the interpolative regime, but they are better at extrapolating (i.e., they are more transferable).  Following these observations, it may be possible to derive a strategy for adjusting the combined weights to achieve the best accuracy on a variety of testing sets.  Although we do not explore such a strategy in this work, we do comment further on the interplay between these two contributions in Section~\ref{sub:showcase}.

\subsection{Comparison with OQML}
\label{sub:comparison-oqml}

It is interesting to compare the performance of our models to that of the operator quantum machine learning (OQML) scheme in Ref.~\citenum{Christensen2019}.
In OQML, a formal dependence on an applied electric field is included in the definition of the (scalar) kernel by assigning fictitious charges to each atom. 
This makes it possible to define derivatives of the kernels relative to an applied field that are naturally covariant and serve as a basis to fit molecular dipoles. 
It should be stressed that, even though the scheme relies on formal atomic charges, it amounts effectively to learning local dipoles, and is therefore similar to our vector model. Whereas in OQML, the energy and dipole regression models are coupled through a scalar constant, our approach allows every property can be trained independently. 

As can be seen in Figure~\ref{fig:lcs_qm9}, the QM7b vector model (the most transferable of the QM7b models) outperforms the FCHL* OQML model by approximately 20\%{}. This is particularly remarkable, because the OQML model of Ref.~\citenum{Christensen2019} was trained on 5000 structures from QM9; the QM7b models, on the other hand, are trained on smaller structures, and are therefore functioning in the much more challenging extrapolative regime.  This is in contrast to the QM9 scalar, vector, and combined models, which are functioning in the interpolative regime in this test.  Here, we observe that the slopes of the QM9-QM9 learning curves are approximately the same as that of the FCHL* (response) curve, but that they have a large offset. In other words, the MuML models achieve an MAE of about \sfrac{1}{3} that of FCHL* with the same amount of data.

\begin{figure*}[hp]
    \centering
    \includegraphics[width=\textwidth]{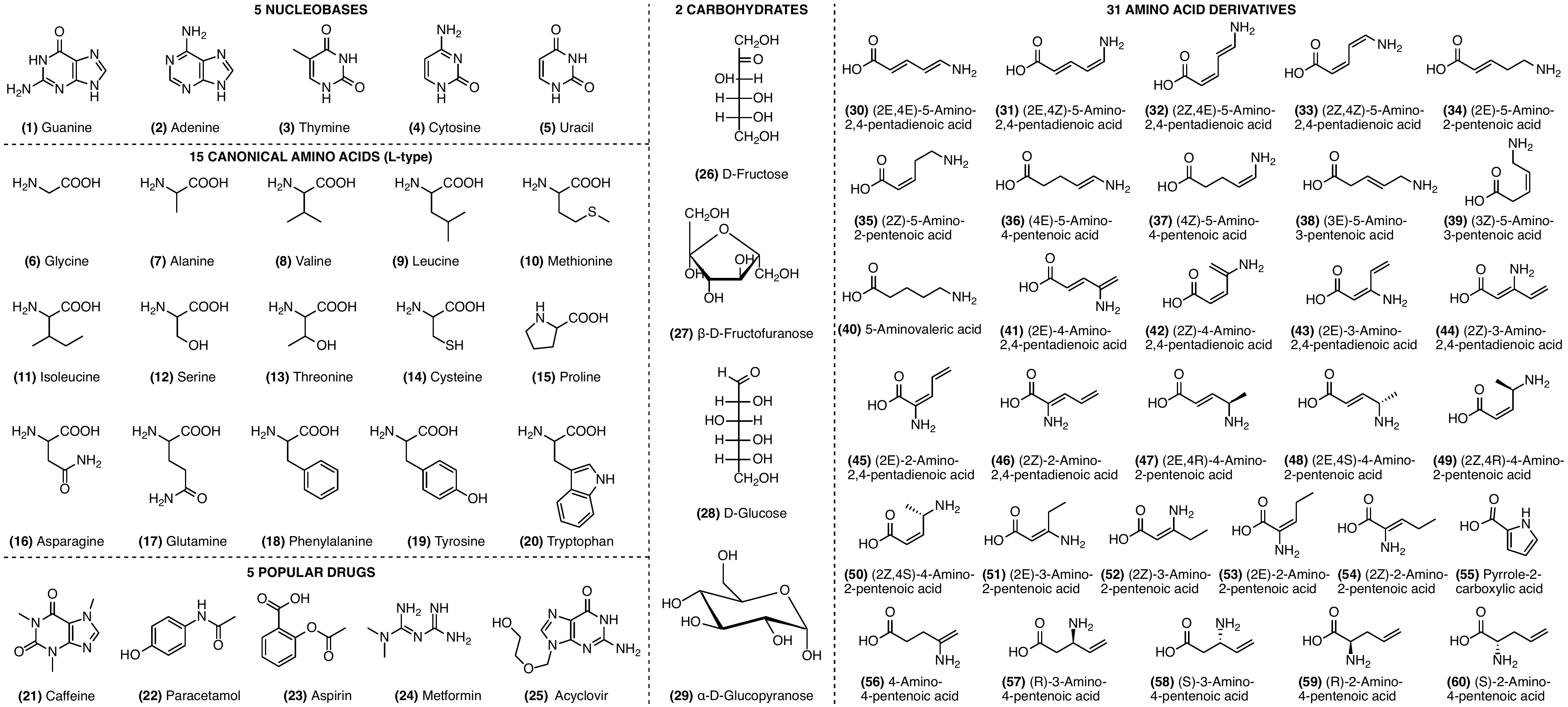}
    \caption{List of molecules included in the MuML showcase dataset. The numerical key is used to identify the various compounds in other figures.}
    \label{fig:muml_showcase_molecules}
\end{figure*}
 
 \begin{figure*}[hp]
    \centering
    \includegraphics[width=\textwidth]{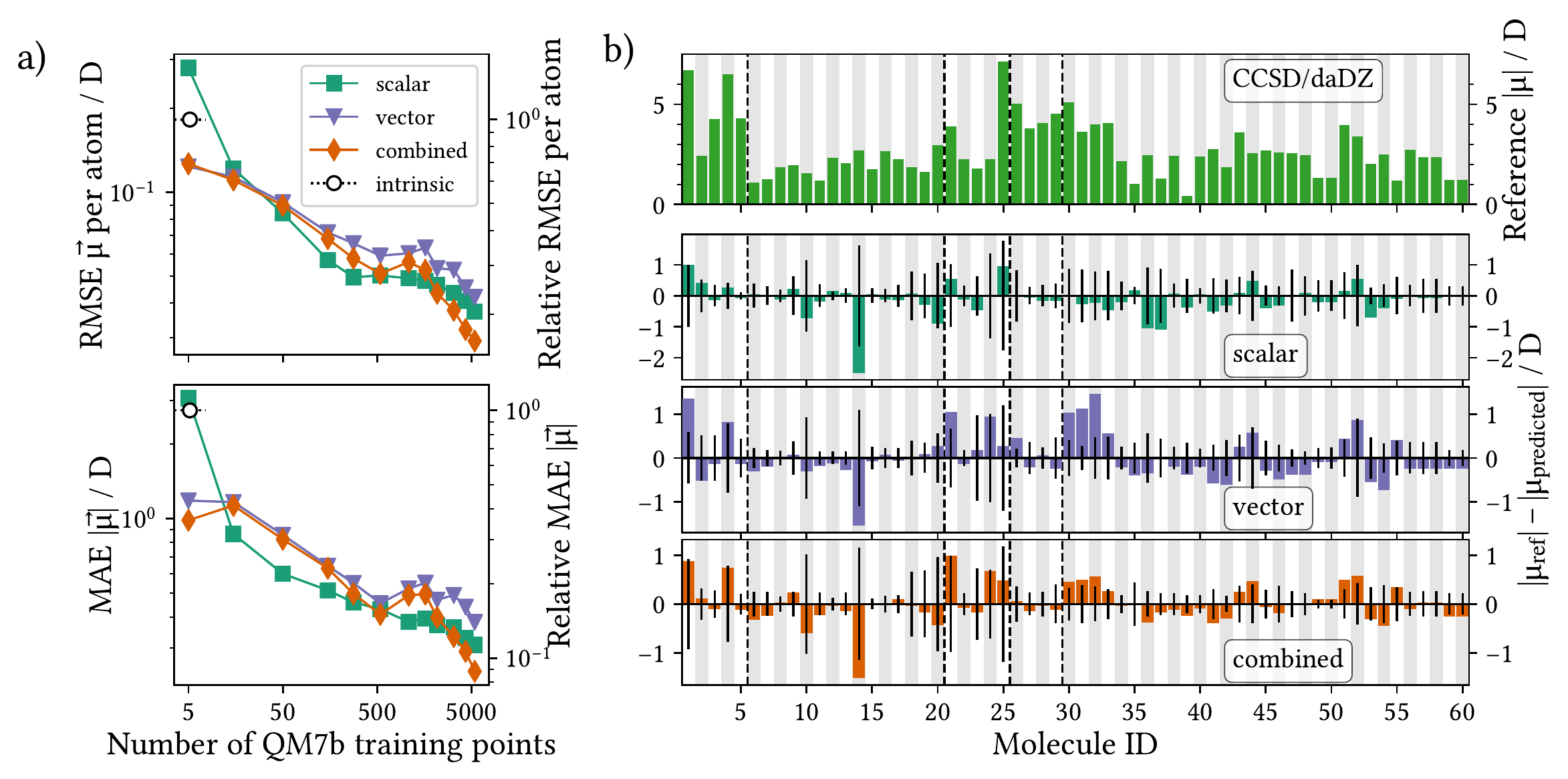}
    \caption{
    Performance of the MuML models on the MuML showcase dataset: a) learning curves by per-atom RMSE (top) and MAE (bottom), b) per-molecule breakdown of the MuML models trained on the full QM7b training set (\num{5400} molecules): norms of the reference dipole moments computed with CCSD/daDZ on the MuML showcase dataset (top) and 
    errors in the norms of the dipole moment predictions across the same set (bottom three). Prediction errors are shown along with error bars from an ensemble of models trained on subsets of the full training set.\cite{musilFastAccurate2019}  The molecule ID is in reference to Figure~\ref{fig:muml_showcase_molecules}.}
    \label{fig:showcase_lc_norm_errs}
\end{figure*}

\subsection{MuML showcase dataset}
\label{sub:showcase}

Similar to Ref.~\citenum{wilkins2019}, we now turn from standard, systematically generated benchmark datasets to a showcase dataset in which chemically relevant molecules have been specifically chosen to test the sensitivity of the ML model to subtler variations in chemical structure and bonding. To this end, we assembled the so-called MuML showcase dataset, which is depicted in Fig.~\ref{fig:muml_showcase_molecules} and comprised of the first 29 molecules of the AlphaML showcase dataset~\cite{yangQuantumMechanical2019,yangPolDataset2019} (and includes the nucleobases, amino acids, sugars, and common drug molecules). The \ce{C8H$_n$} isomers from the AlphaML showcase dataset were discarded (because they all have very small dipole moments), and substituted with 31 \ce{C4H$_n$NH_2COOH} amino acid derivatives, with dipole moments spanning a broad range from \SIrange{0.5}{6}{\debye}. Molecular geometries and dipole moments for these new molecules were obtained using the same protocol described in Refs.~\citenum{wilkins2019,yangQuantumMechanical2019}. For reference, the dipole moment norms computed with CCSD/daDZ on the MuML showcase are shown in the top panel of Figure~\ref{fig:showcase_lc_norm_errs}b.

The learning curves of the three dipole models on the MuML showcase dataset are shown in Figure~\ref{fig:showcase_lc_norm_errs}a.  All three models achieve an accuracy comparable, in absolute terms, to that on QM9.  The (unadjusted) combined model narrowly outperforms both the scalar and vector models.
Even in this extrapolative regime, the accuracy of MuML is competitive with that of B3LYP: for the largest training set size, MuML achieves errors ($\mathrm{RMSE} = {}$\SI{0.029}{\debye} per atom, $\mathrm{MAE} = \SI{0.24}{\debye}$) that are only \SI{30}{\percent} larger in RMSE (\SI{56}{\percent} larger in MAE) than those of B3LYP relative to CCSD ($\mathrm{RMSE} = \SI{0.019}{\debye}$ per atom, $\mathrm{MAE} = \SI{0.19}{\debye}$). The dramatic increase in accuracy observed when training on the larger QM9 molecules (see Fig.~\ref{fig:lcs_qm9}) suggests that it is possible to train a MuML model that will outperform DFT on this showcase dataset. Unfortunately, the cost of performing LR-CCSD calculations on thousands of QM9-sized molecules is still prohibitive at the current time.

Due to the relatively small number of molecules in the MuML showcase set, we can examine the performance of the MuML models for each of the molecules individually. 
Furthermore, we can also benchmark the uncertainty quantification scheme discussed in Section~\ref{sec:uq}. Each of the eight models in the committee model was trained on a sample of $\num{2700}$ molecules (\SI{50}{\percent}) of the full training set), drawn from the full QM7b training set without replacement. The calibrated error predictions were then validated against the QM7b test set; additional details can be found in the \theSI{}.  The overall errors of the ensemble averages are comparable to (if slightly higher than) those of the model trained on all $\num{5400}$ points.

\begin{figure*}
    \centering
    \includegraphics{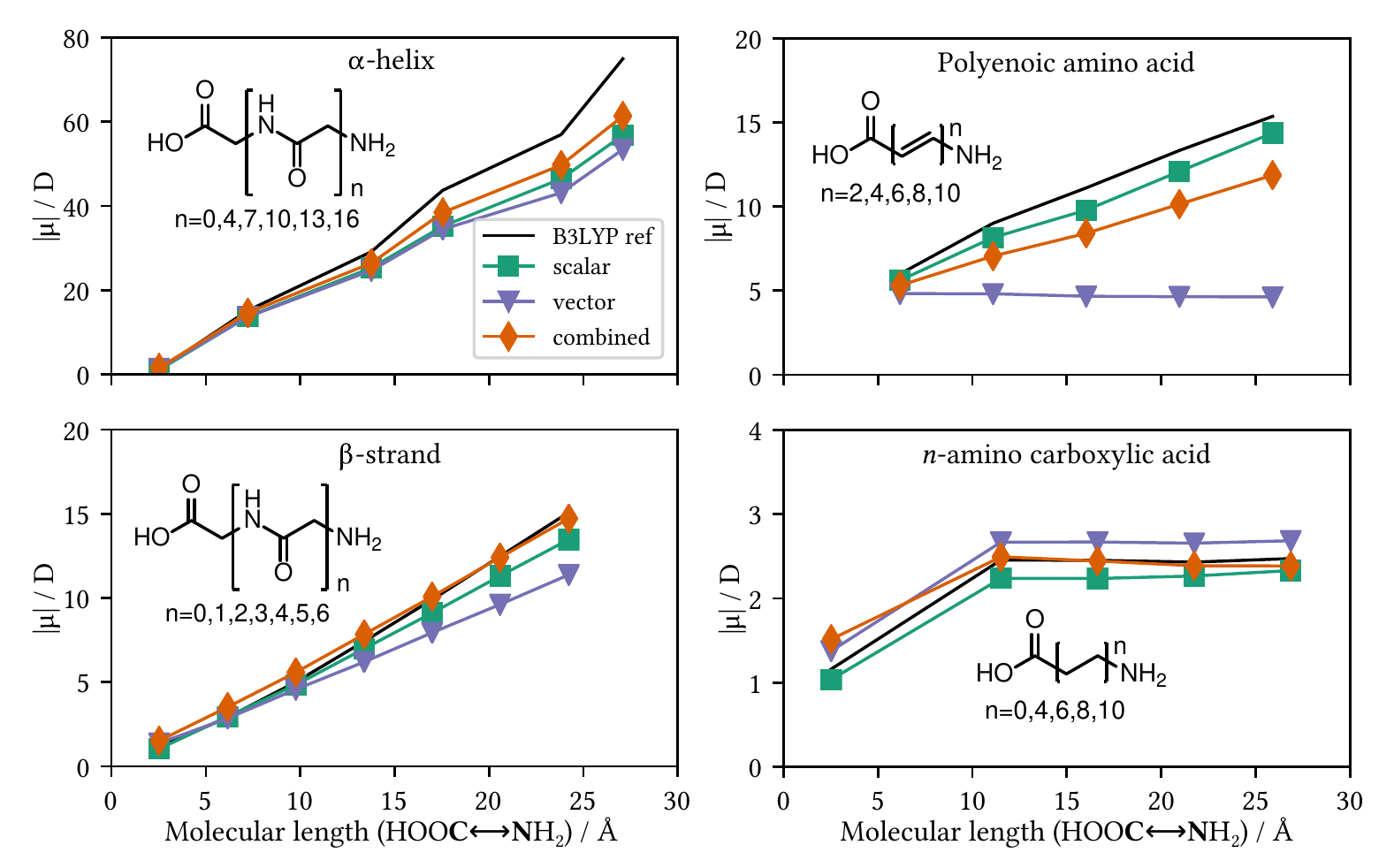}
    \caption{Dipole moment predictions for the four ``challenge'' cases: polyglycine in the $\alpha$-helix conformation, polyglycine in the $\beta$-strand conformation, polyenoic amino acids (\textit{trans}-polyacetylene-bridged glycine), and $n$-amino carboxylic acids (polyethylene-bridged glycine).  All three models perform fairly well on the polypeptides, where the charge polarization is a mostly local phenomenon.  On the unsaturated bridged glycine, however, the vector model completely fails, with only the scalar model maintaining accuracy (and the combined model suffering from the inclusion of the unphysical asymptotic behavior of the vector model).  The saturated bridged glycine has completely different behaviour, with the dipole saturating to a small constant value; all three models predict this trend accurately.  All predictions are from the MuML models trained on $\num{5400}$ QM7b molecules (not the ensemble models).} %
    \label{fig:challenge_comparison}
\end{figure*}

Figure~\ref{fig:showcase_lc_norm_errs}b shows the breakdown of the errors of the ensemble average, along with the uncertainties predicted from the ensemble.  Note that the errors are shown reversed from the usual convention--they are shown as reference minus predicted--and the error bars are shown centered about zero. 
Both the predicted uncertainties and the ensemble-average residuals show no apparent systematic patterns across this set of molecules, although there are some outliers.  All three MuML models perform particularly poorly on Molecule 14 (cysteine), and the uncertainty estimate is also relatively high for this molecule.   The evidence suggests that the high errors and uncertainties are a consequence of the highly-polarizable nature of sulfur, given that the models also give large overpredictions and high uncertainties in the case of methionine, the only other S-containing molecule in the MuML showcase dataset.  Other relatively large errors (and large uncertainties) are seen on all models for Molecule 1 (guanine), Molecule 21 (caffeine), Molecule 24 (metformin), and Molecule 25 (acyclovir); the vector and combined models additionally give large errors and uncertainties for Molecule 4 (cytosine) and Molecule 23 (aspirin).

Overall, the prediction errors are consistent with the error bars, with \SI{88}{\percent} of the scalar predictions, \SI{55}{\percent} of the vector predictions, and \SI{72}{\percent} of the combined predictions falling within one error bar of the reference (compare this with the \SI{68}{\percent} expected if the prediction errors were to follow a Gaussian distribution with a standard deviation equal to the error bar). Thus, the uncertainty quantification scheme applied herein provides a reliable estimate of the model accuracy, improving our interpretation of the model results in the extrapolative regime where the errors can be several times larger than those in the original testing set.

The only cases in which the predictions are farther than two error bars from the reference is that of Molecules 30, 31, and 32: these show large errors but small uncertainties in the vector model.  Together with the similar structure of these molecules--they are all polyenoic amino acids, effectively an amine group and a carboxylic acid joined by a fully conjugated polyacetylene chain/linker--these deviations suggest a \emph{systematic} error in the vector model predictions.  The delocalized nature of the conjugated chains in these molecules suggests that the error could derive from a non-local effect that the vector model, with its finite cutoff and strictly local environmental dependence of the atomic dipoles, fails to capture. 

In order to provide a more systematic, and far more stringent, test of our models' extrapolative capabilities, as well as to investigate the effect of non-local effects on each of the models, we designed four new ``challenge'' test sets, each of which consists of a series of approximately linear (pseudo-1D) molecules with polar groups and (in three of the four sets) large separations of charge, thereby giving rise to large dipole moments.
More specifically, we considered polymers of the glycine amino acid, in both the $\alpha$-helix and $\beta$-strand configurations, as well as a series of polyenoic amino acids, with an amine group and a carboxylic acid group separated by a polyacetylene spacer.  Finally, a set of $n$-amino carboxylic acids (the saturated analogs of the polyenoic amino acids) was included to investigate the effect of saturation in the spacer on the molecular dipole moments and the model predictions.  Because of the large size of these molecules (up to 122 atoms, of which 69 were heavy/non-H atoms for the longest $\alpha$ helical configuration), we used B3LYP/daDZ references and models.
Figure~\ref{fig:challenge_comparison} contrasts the growth of the dipole with chain length with the predictions of the scalar, vector, and combined models. In the case of polyglycine, the three models capture at least qualitatively the trend, with the vector model usually under-predicting the slope, and the combined model performing substantially better than either the scalar or the vector model.
In the case of the polyenoic amino acids, however, the vector model breaks down completely, predicting a constant dipole as a function of chain length. The scalar model most closely approaches the correct slope, and the combined model shows the correct trend, but with a smaller slope to the pure scalar model.
The saturated $n$-amino carboxylic acids showed a completely different trend, with the total dipole levelling off to a constant small value and the model predictions essentially following this trend.  This contrast points to the conjugated, non-local nature of the polyenoic acids as an essential ingredient to their large dipole moments: indeed, their saturated counterparts have stronger charge locality and cannot transfer/delocalize charge across the whole molecule, like the unsaturated chains can.

\begin{figure*}[p]
    \centering
    \includegraphics[width=1.0\textwidth]{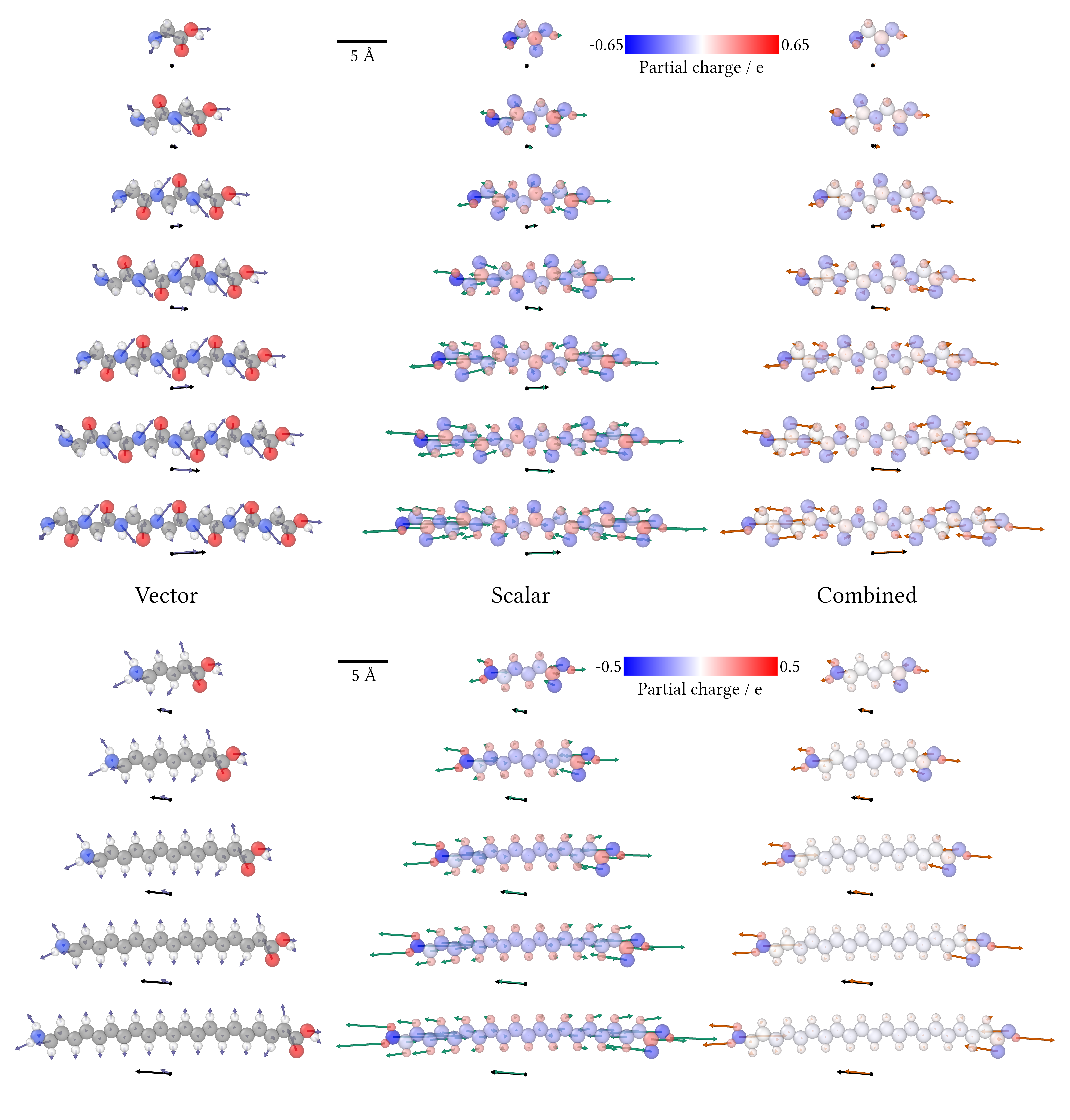}
    \caption{
A representation of the per-atom contributions to the total dipole for two of the challenge systems: polyglycine in the $\beta$-strand conformation (top), and polyenoic amino acids in the all-trans conformation (bottom). 
Vector per-atom dipoles are defined in Eq.~\eqref{eq:vector-peratom-dipoles}, and plotted exaggerated by a factor of 5 for visibility; atoms are colored according to the atom type. 
Scalar per-atom dipoles are defined as the partial charges multiplied with the displacement vectors (referenced to the molecule's center of geometry), as in Eq.~\eqref{eq:partial-charge-atomsum}; atomic charges are also represented as atom colors according to the displayed color scale. 
Per-atom dipoles for the combined model are the (appropriately weighted) sums of the respective scalar and vector per-atom dipole predictions.  
The total dipoles, in each case equal to the sum of the per-atom predictions, are shown below each molecule along with (in black) the reference dipole moment computed from B3LYP/daDZ.  The per-atom arrows for the vector model are exaggerated by a factor of 5 for visibility.  The scale bar shows the maximum range of sensitivity (\SI{5}{\angstrom}) of the partial charges and atomic dipoles to their environments.
    Visualizations created with Ovito.~\cite{stukowskiVisualizationAnalysis2009}
    }
    \label{fig:peratom_showcase_mu}
\end{figure*}

To gain deeper insight into the performance of the different MuML models as well as the physical effects that determine the breakdown of the vector model, we computed atomic contributions to the dipole moment--both the vector predictions and the partial charges (for the models that use them) for each atom--and represented them together with the molecular structure in Figure~\ref{fig:peratom_showcase_mu}.  Here, we discuss only the $\beta$-strands and the polyenoic amino acids, as the observations for the $\alpha$-helical structures are very similar to those for the $\beta$-strands.  The per-atom breakdown for the $\alpha$-helices  and the $n$-amino carboxylic acids can be found in the SI.
In the case of the polyglycine $\beta$-strand, each monomeric unit is polar.  Since the total dipole is almost entirely made up of these local monomeric contributions, the vector model based on local atomic dipoles captures the correct scaling behavior with system size. The scalar model also captures the correct behavior, as each molecular unit is (approximately) neutral and contributes a roughly constant term (even though individual atomic dipoles grow larger for atoms that are farther away from the molecular center). The $n$-amino carboxylic acid also exhibits strongly localized physics, with the molecular dipole moment being mostly generated by local polarization of the end groups, and all three MuML models are able to give accurate predictions.
Clearer differences between the MuML models arise in the case of the polyenoic amino acids. The non-polar spacer is (correctly) predicted to contribute very little to the total dipole, while the amine and carboxylic acid functional groups each bear a (roughly) constant dipole, which results in a prediction that is independent on the length of the spacer. The scalar model, on the other hand, predicts net positive and negative scalar charges on the amine and carboxylic acid groups, and as a consequence predicts a total dipole that scales linearly with the length of the spacer. 
Even if it underpredicts the total dipole, the combined model most closely reflects conventional chemical wisdom: it predicts negligible charges along the polyacetylene spacer with only the polar end groups contributing to the total dipole. Since the end groups carry a net positive and negative charge, the total dipole increases with their separation.

These observations reflect the shortcomings of a local ML model, similar to what was observed in Ref.~\citenum{wilkins2019} for the molecular polarizability of conjugated hydrocarbons (e.g., alkenes and acenes). SOAP features are computed with a cutoff of 5~\AA{}, and cannot therefore describe structure-property correlations beyond this limit. The scalar model circumvents this limitation by assuming that atomic charges are local, and that the non-locality of the dipole moment is entirely captured by the spatial separation of the atomic charges. 
As shown in the SI, the radial scaling functions of the two models, which reflect how quickly the influence of far-away atoms decay, is consistent with the greater non-locality of the vector model. The radial scaling of the scalar models decays rapidly, well before the neighbor list cutoff, while that of the vector model indicates that correlations beyond \SI{5}{\angstrom} would be needed to describe molecular dipoles as a sum of local contributions. 

Finally, while the local partial charges and dipoles provided by this analysis bear some similarities to the electron density decomposition schemes discussed in Section~\ref{sec:theory}, they should not be confused.  The partitioning scheme shown here does not use the electron density; rather, it provides an interpretable description of how the ML model arrives at its prediction of the total dipole, allowing us to verify whether or not it includes the appropriate physics.

\section{Conclusions}

In this work, we have introduced a set of models for predicting molecular dipole moments that we collectively refer to as ``MuML''. These models rely on a local, atom-centered description of molecular structure that fulfills the symmetries of the target property. We compare a vector model that predicts atom-centered dipolar contributions with a scalar model that predicts atomic charges entering into a physics-based expression for the contribution to the total dipole moment. 
Training on reference CCSD calculations performed on a set of small organic molecules, both models can achieve a similar accuracy of around \SI{0.1}{\debye}, which is comparable  to the accuracy of DFT, with a slight improvement made possible by combining the two models. 
The differences between the models are more noticeable---up to \SI{40}{\percent} RMSE--when considering the transferability to larger molecules, namely the QM9 dataset. Here, the vector model seems to be more robust while the scalar model appears to overfit, with a model trained on 5000 small molecules giving worse performance than one trained on only 500. 
Even with these limitations, the vector model outperforms  a state-of-the-art model based on the FCHL* framework~\cite{Christensen2019}, even though FCHL* model is trained on another subset of QM9 molecules, and is therefore operating in the interpolative regime.
When we use training structures from QM9, the performance of MuML dramatically improves, and we observe a three-fold reduction in the error. 
The accuracy of the combined model can be improved by adjusting the relative weight of the scalar and vector models, with better interpolative performance observed in the limit of large scalar weights, and better extrapolative performance when using large vector weights. 
State-of-the-art performance for MuML is also observed for a showcase dataset of even larger and more complex molecules, where the scalar model shows improved performance relative to the vector model, and the combined model approaches the accuracy of DFT. For these molecules, we also show that a calibrated committee model can accurately estimate the uncertainty in the model predictions, thereby further improving the reliability of predictions in this challenging extrapolative regime.  

In this work, we finally pushed MuML to its breaking point by performing predictions on a set of polymers of increasing length, that extend far beyond the cutoff radius of the atom-centered features used to describe the molecules. 
In this regime, the vector model can predict reasonably well the molecular dipole moment of polyglycine, for which each monomeric unit contributes a dipolar term.  It fails dramatically, however, for the polyenoic amino acid series, where the increase in the molecular dipole moment arises because of charge separation by fully conjugated (but non-polar) spacer units.
The scalar model, on the other hand, recovers this effect correctly because the geometric separation between atoms is built into the form of the kernel, introducing an element of non-locality. 

The combination of these two models makes it possible to improve the performance of MuML, even though the optimal combination of weights depends rather strongly on the nature of the test molecules. This suggests that, even when taken together, local vector and scalar models of the dipole only partially capture the physics of polarization, affecting the overall model's transferability. An explicit treatment of long-range effects using a charge equilibration scheme~\cite{Ghasemi2015}, or incorporating long-range correlations by long-distance equivariant features~\cite{grisafiIncorporatingLongrange2019}, might further improve the accuracy of MuML, which is already competitive with that of hybrid DFT calculations while being dramatically less computationally expensive.

Another direction for further research involves the modelling of condensed phases. The presence of periodic boundary conditions makes the position operator ill-defined. As a consequence, an expression like Eq.\eqref{eq:dipole-charges} cannot be used to define polarization in the condensed phase, which makes the scalar and, by extension, the combined models inapplicable.  One way around this limitation is to instead model the position of Wannier function centers, so that each point in the unit cell is an integer multiple of the electron charge, thereby preserving the lattice condition for polarization in a periodic medium (see e.g. \citet{Spaldin2012} or \citet{Resta2018}).  Current implementations of the idea, however, predict the position of centers attached to an atom~\cite{Zhang2019}, so that the framework is effectively equivalent to learning atom-centred dipoles.  
Indeed, a vector model can be readily applied to bulk systems, and has already been used successfully to predict the infrared spectrum of liquid water~\cite{kapilInexpensiveModeling2020}. It is not obvious, however, that this methodology will work well in systems where there is significant delocalization of charge.  
Incorporating ideas from the modern theory of polarization~\cite{Resta2018}, learning the Born effective charge tensors, or taking a more decidedly data-driven approach by using long-range features without explicitly incorporating a physical description of electrostatics all provide possible strategies to apply to condensed phases a model that can capture, like MuML, the different phenomena that give rise to permanent or transient polarization. 

\section*{\theSI{}}
The supplementary material contains further details about the derivation, implementation and benchmarks of the method, including:
\begin{itemize*}
    \item Convergence of the scalar and vector models on QM7b
    \item Kernel optimization procedure
    \item Radial scaling function for scalar and vector models
    \item Uncertainty quantification calibration procedure
    \item QM7b Learning curves for B3LYP dipoles
    \item Comparison of B3LYP, CCSD, and SCAN0 dipole moment predictions on the MuML showcase
    \item Per-atom breakdown of the alpha-helix and $n$-amino carboxylic acid predictions
\end{itemize*}

\section*{Acknowledgments}

D.M.W. was supported by the European Research Council under the European Union's Horizon 2020 research and innovation programme (grant agreement no. 677013-HBMAP), and benefited from generous allocation of computer time by CSCS, under project ID s843.  M.V. and M.C. were supported by the Samsung Advanced Institute of Technology, and by the NCCR MARVEL of the Swiss National Science Foundation (SNSF).  Y.Y. and R.A.D. acknowledge support from Cornell University through start-up funding. This research used resources of the National Energy Research Scientific Computing Center (NERSC), a U.S. Department of Energy Office of Science User Facility operated under Contract No. DE-AC02-05CH11231.

\section*{Data availability}

The geometries and data used to fit the models in this paper are publicly available\cite{veitQuantumMechanical2020} in Materials Cloud under the DOI \url{https://doi.org/10.24435/materialscloud:2k-3h}; the models themselves are available\cite{veitModelsSource2020} at Zenodo under the DOI \url{https://doi.org/10.5281/zenodo.3820297}.  The code used to fit and evaluate the models is publicly available at \url{https://github.com/max-veit/velociraptor}.

\include{si-pdf}

\end{document}

\title{Supplementary Information\\Predicting molecular dipole moments by combining atomic partial charges and atomic dipoles}

\author{Max Veit}
\affiliation{Laboratory of Computational Science and Modeling, IMX, \'Ecole Polytechnique F\'ed\'erale de Lausanne, 1015 Lausanne, Switzerland}

\author{David M. Wilkins}
\affiliation{Laboratory of Computational Science and Modeling, IMX, \'Ecole Polytechnique F\'ed\'erale de Lausanne, 1015 Lausanne, Switzerland}

\author{Yang Yang}
\affiliation{Department of Chemistry and Chemical Biology, Cornell University, Ithaca, NY 14853, USA}

\author{Robert A. DiStasio Jr.}
\affiliation{Department of Chemistry and Chemical Biology, Cornell University, Ithaca, NY 14853, USA}

\author{Michele Ceriotti}
\email{michele.ceriotti@epfl.ch}
\affiliation{Laboratory of Computational Science and Modeling, IMX, \'Ecole Polytechnique F\'ed\'erale de Lausanne, 1015 Lausanne, Switzerland}

\maketitle

\newpage

\listoffigures

\section{Kernel convergence}

\begin{figure}
    \centering
    \includegraphics{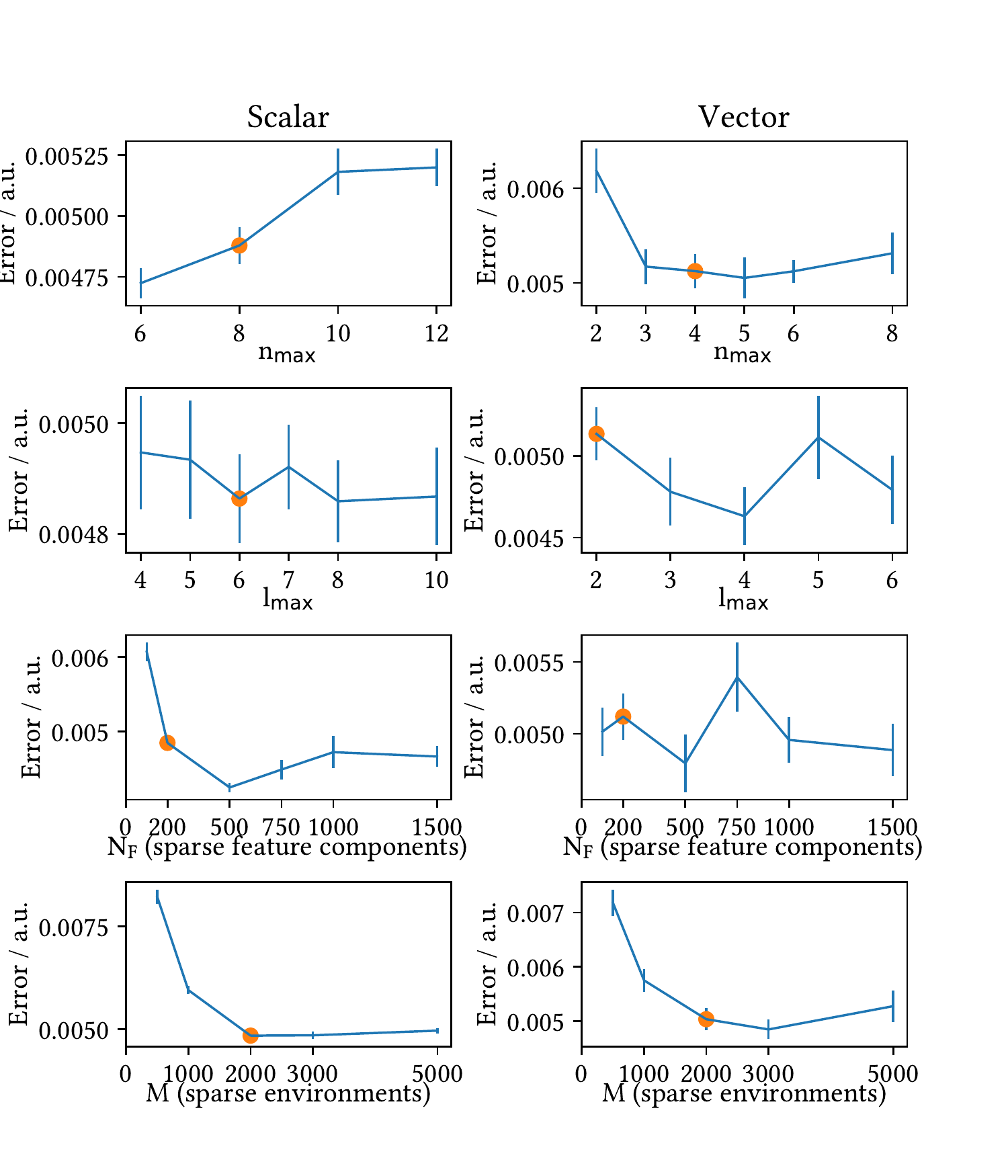}
    \caption[Kernel convergence]{Convergence of the scalar and vector kernels as four parameters are varied.  The error for the scalar kernels is computed over a six-fold cross-validation partitioning randomly selected from the 5400 QM7b training molecules used in the main paper; the vector kernel error is computed over a four-fold CV partitioning taken from the same set.  Each plot shows the dependence of the CV-error as the labelled parameter is varied around the final, chosen values (indicated by orange dots).  The error bars show the standard deviation of the cross-validation error.  All other kernel parameters were set to the final optimized values.}
    \label{fig:kernel_convergence}
\end{figure}

The scalar and vector kernels both depend on several discrete parameters that affect their predictive performance. These are the radial ($n_\text{max}$) and angular ($l_\text{max}$) band limits of the spherical expansion; the number of sparse feature components ($N_F$) selected, and the number of sparse environments ($M$) selected for the fit.  Of these, $n_\text{max}$, $l_\text{max}$, and $N_F$  affect the speed of kernel building, and hence prediction. $M$, on the other hand, is the size of the matrix to be inverted; therefore, the time of the fitting (weight-computing) step depends cubically on $M$.  The prediction time also depends on $M$, but only linearly.

Figure~\ref{fig:kernel_convergence} shows the dependence of the fitting error on these four parameters for both the scalar and vector kernels.  (For the vector kernel, the $\lambda=0$ power spectrum is constrained to use the same parameters as the $\lambda=1$ power spectrum, to limit the size of the optimization space.)  Both $n_\text{max}$ and especially $l_\text{max}$ are too noisy to show any clear dependence, so sensible standard values were chosen (marked in orange in the plots) -- the vector kernel values being lower due to the increased computational cost of computing this kernel.  The $N_F$ dependence is pronounced only for the scalar kernel, so the value of 200 was chosen to give reasonable performance for both kernels.  Finally, the number of sparse environments $M$ is the only parameter to have a real, pronounced effect on the final CV-error; the value of 2000 was chosen for both as a good compromise between lowering the final error and keeping the computational cost within bounds.

\section{Kernel optimization}

The remaining, continuous parameters are optimized, again on the 5400 training molecules of QM7b but this time with four-fold CV for both scalar and vector kernels.  The optimization consists of two layers: an outer Nelder-Mead (simplex) optimization of the atom width ($\theta_w$) and the radial scaling parameters $r_0$ and $m$, initialized with the simplex in Table~\ref{tab:kinit_simplex}, coupled with an inner optimization of the regularization parameters (which have a much simpler dependence for the final error).  For the vector kernels, where only one regularization parameter is used, the inner optimization uses Brent's algorithm for one-dimensional optimization problems (the default in SciPy 1.4.1).  For the scalar kernels there are two regularization parameters, which are optimized again using the Nelder-Mead method.  Both regularizer optimizations are performed in log-space to follow the typical logarithmic dependence of the results on the regularizers.  All Nelder-Mead optimizations are terminated once the absolute value of change of the parameter vector is less than 0.01.

\begin{table}[ht]
    \centering
    \begin{tabular}{r c c c}
    Point & $\theta_w$ &    $r_0$   &      $m$ \\ \hline
    0  & 0.3    &    3.0    &    5.0 \\
    1  & 0.3    &    4.0    &    5.0 \\
    2  & 0.3    &    2.0    &    3.0 \\
    3  & 0.5    &    3.0    &    6.0
    \end{tabular}
    \caption{Initial simplex for the parameters of the outer kernel optimization}
    \label{tab:kinit_simplex}
\end{table}

\subsection{Radial scaling}
\begin{figure}
    \centering
    \includegraphics{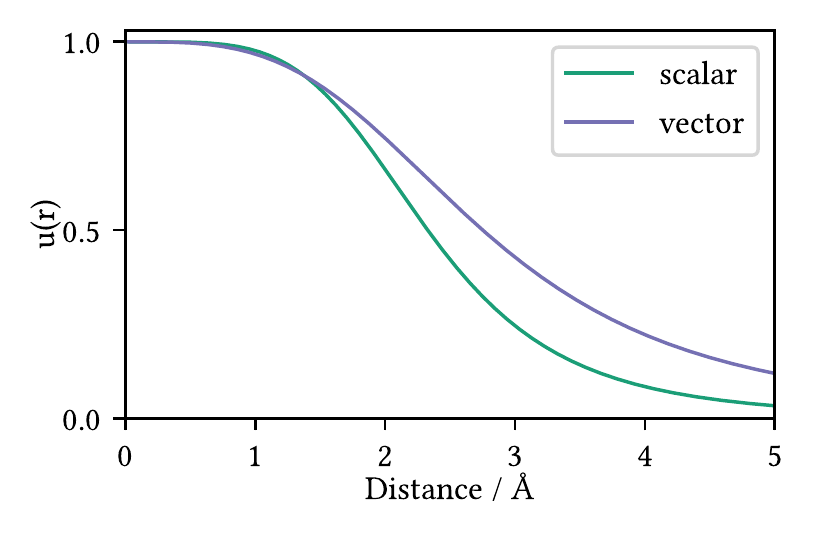}
    \caption[Radial scaling function]{The radial scaling functions found through separate optimizations of the scalar and vector kernels.  The final cutoff at \SI{5}{\angstrom} is sharp.}
    \label{fig:radial_scaling}
\end{figure}

The radial scaling functions\cite{willattFeatureOptimization2018} resulting from the optimized kernel parameters (shown in Table~I of the main paper) are plotted in Figure~\ref{fig:radial_scaling}.

\section{Error estimation}

\begin{figure}
    \centering
    \includegraphics[width=0.3\textwidth]{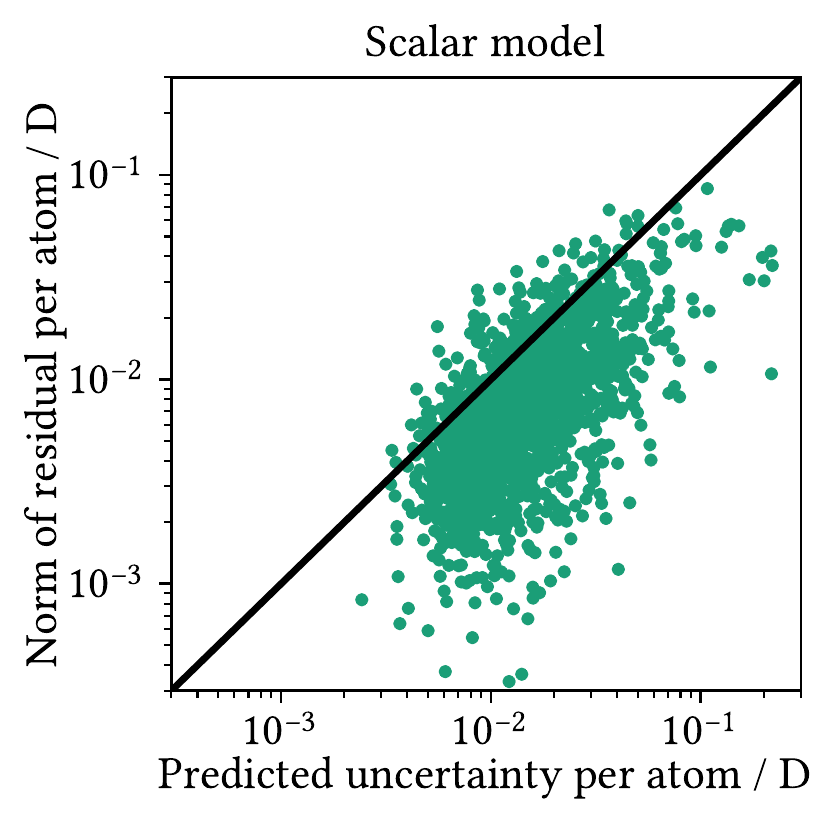}
    \includegraphics[width=0.3\textwidth]{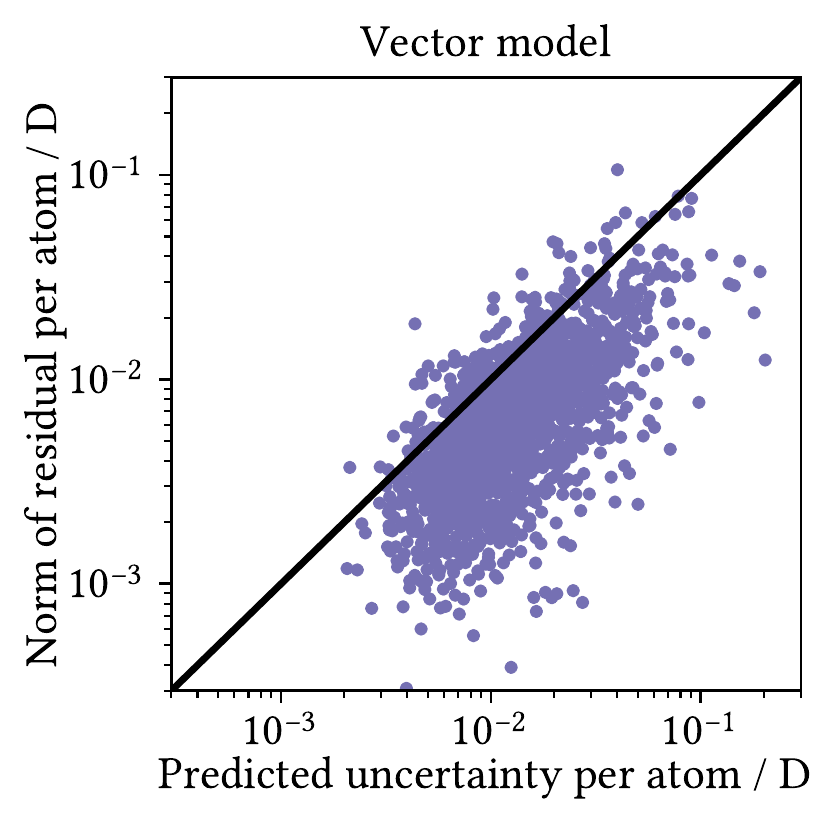}
    \includegraphics[width=0.3\textwidth]{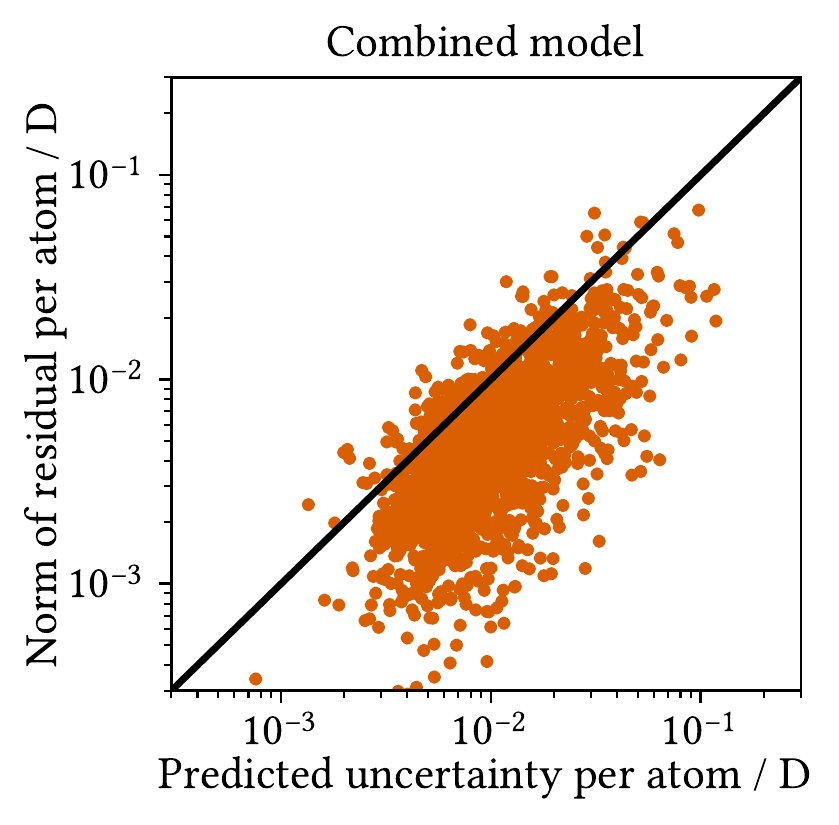}
    \caption[Predicted uncertainties versus model residuals]{Predicted uncertainties versus model residuals for the scalar, vector, and combined models.  The black diagonal lines show $y=x$.}
    \label{fig:error-estimation-qm7-val}
\end{figure}

The error estimation was calibrated using the ``internal validation scheme'' described in Ref.~\citenum{musilFastAccurate2019}: For all training points included in 4 or fewer of the 8 ensemble models, the ratio of the mean residuals of the models \emph{not} including that point divided by the standard deviation of the prediction of the same models was taken; this ratio was averaged in quadrature over all eligible training points and all three Cartesian components to get the overall isotropic scaling factor $\alpha_0$ for that model (scalar, vector, or combined).  The predicted uncertainty for any new data point is then the standard deviation of the ensemble predictions multiplied by the model's scaling factor.  The reliability of these error estimations was tested on the QM7b validation set:  Figure~\ref{fig:error-estimation-qm7-val} plots the predicted (scaled) uncertainty against the residual of the ensemble mean.  All three models show that the actual residual never significantly exceeds the predicted uncertainty, meaning that the uncertainty estimator will never predict a value that is too low, i.e. it will not overestimate the model's confidence in a given prediction.  The actual residual can, of course, be smaller than the predicted uncertainty because it is expected to vary along a distribution whose width is given by the predicted uncertainty value.

\pagebreak[4]

The overall error measures for the ensemble models are shown in Table~\ref{tab:full-ensemble-comparison} below.  The ensemble average produces errors that are consistently slightly larger than those of the single model trained on the full 5400 QM7b molecules.  Therefore, we should prefer the single full model in order to make accurate predictions, whilst using the ensemble model if we require error estimation.

\begin{table}[ht]
\centering
\begin{tabular}{l
        S[table-format=1.6]
        S[table-format=1.6]}
     Error measure &  \text{Full QM7b model error}/\si{\debye} & \text{Ensemble average error}/\si{\debye} \\ \hline
     Scalar RMSE   & 0.0373 & 0.0371 \\
     Scalar MAE    & 0.309  & 0.314 \\
     Vector RMSE   & 0.0421 & 0.0458 \\
     Vector MAE    & 0.3817 & 0.403 \\
     Combined RMSE & 0.0293 & 0.0321 \\
     Combined MAE  & 0.242  & 0.274
\end{tabular}
\caption{Comparison of error measures for the models trained on the full set of 5400 QM7b molecules with the average of ensemble of models trained on samples of the same set}
\label{tab:full-ensemble-comparison}
\end{table}

\clearpage

\section{Other figures}

\begin{figure}[ht]
    \centering
    \includegraphics{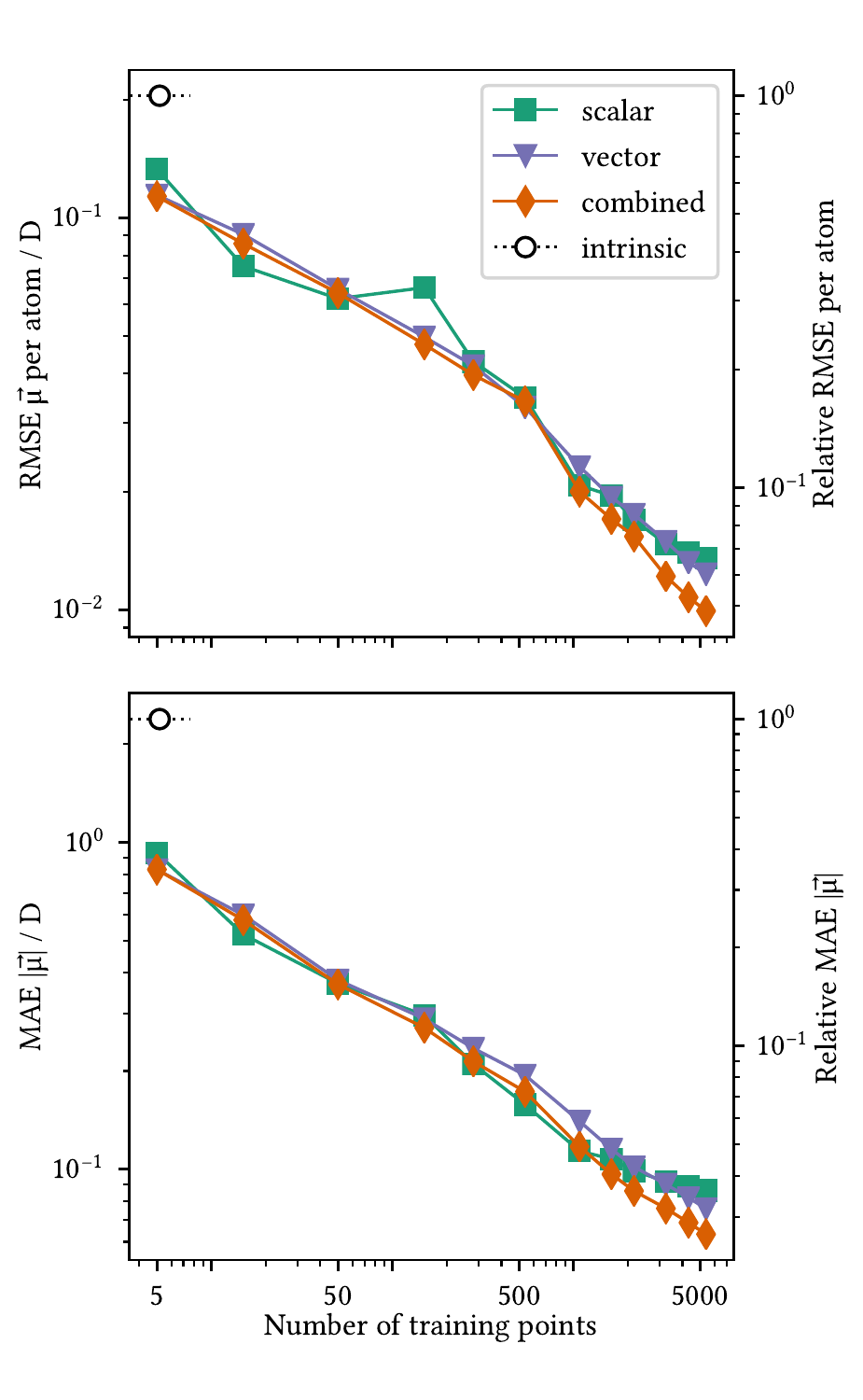}
    \caption[B3LYP learning curves on QM7b]{Learning curves on QM7b, analogous to Figure~1 of the main paper but using B3LYP/daDZ\cite{beckeDensityFunctional1993,kimComparisonDensity1994,woonGaussianBasis1994} dipole data instead.  Note the results are very similar, with the B3LYP fits being slightly less accurate than the CCSD ones.}
    \label{fig:lc_qm7_b3lyp}
\end{figure}

\begin{figure}[hp]
    \centering
    \includegraphics{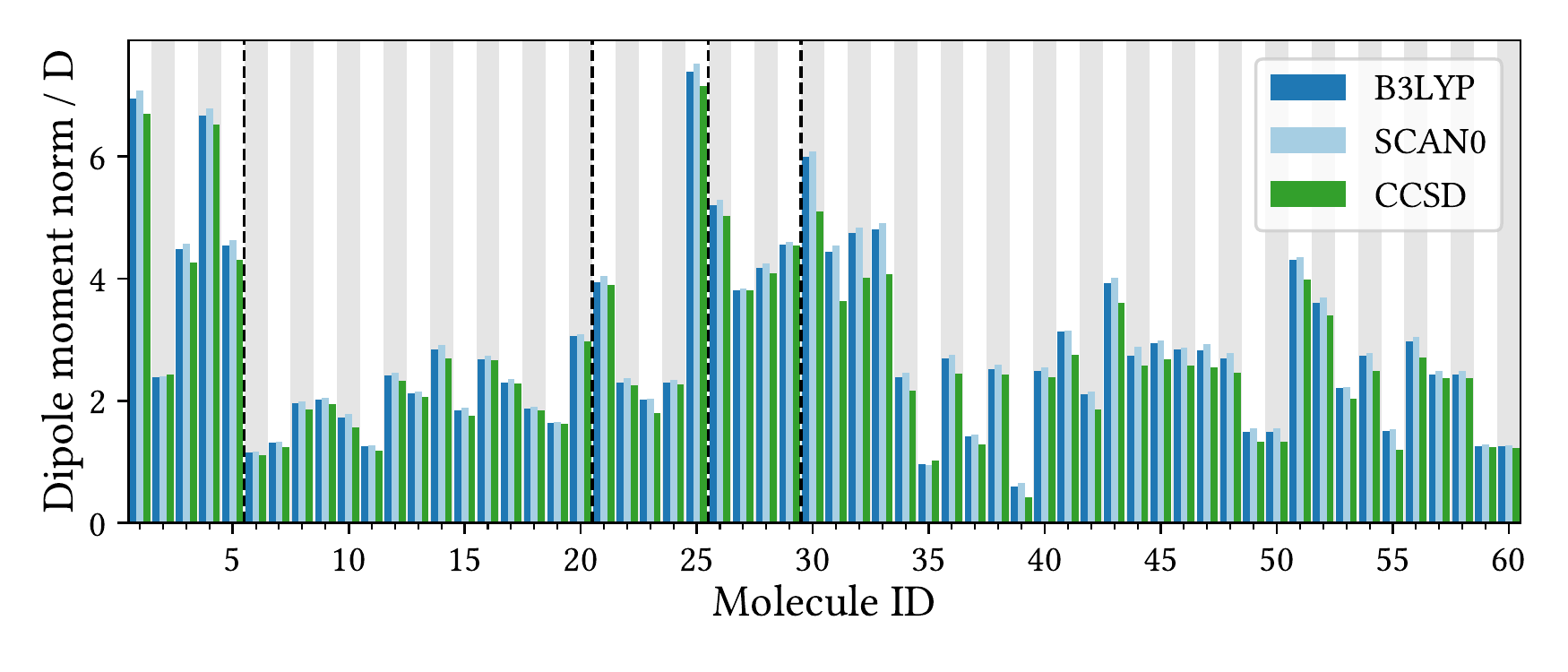}
    \caption[Comparison of MuML showcase dipoles for different levels of theory]{Comparison of the norms of the dipole moments computed on the MuML showcase set by three different methods: B3LYP,\cite{beckeDensityFunctional1993,kimComparisonDensity1994} SCAN0,\cite{huiSCANbasedHybrid2016} and CCSD.\cite{kochCoupledCluster1990}  Note the norms are largely consistent, with the exception of the conjugated amino acids (Molecules 30-33) where the DFT norms are up to \SI{17}{\percent} larger than the CCSD ones.}
    \label{fig:muml_norms_comparison}
\end{figure}

\begin{figure}[hp]
    \centering
    \includegraphics[width=\textwidth]{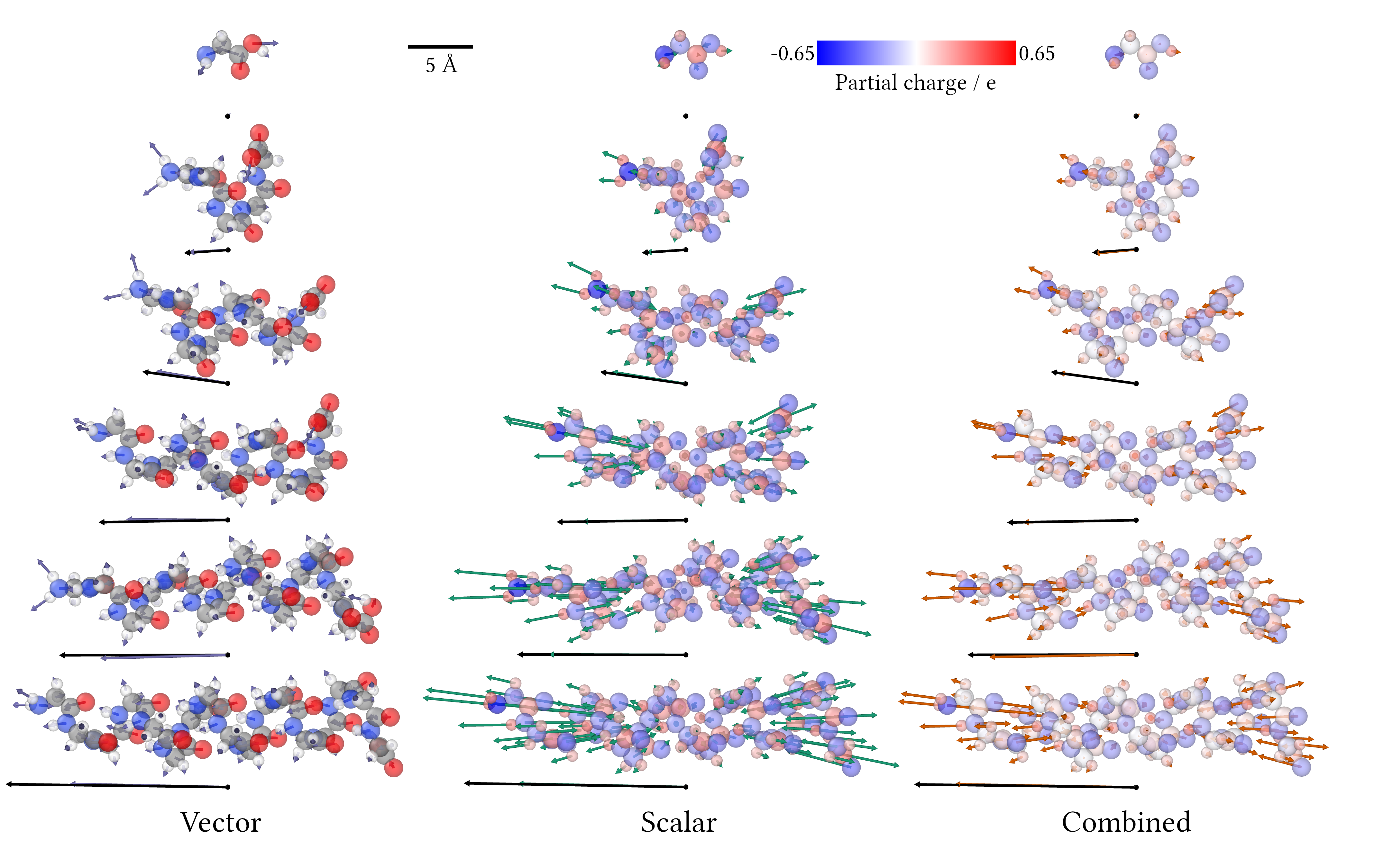}
    \caption[Per-atom breakdown of $\alpha$-helix predictions]{Per-atom breakdown of the MuML predictions on the $\alpha$-helix challenge series.  The results are broadly similar to the $\beta$-strands, but with much larger dipoles (in part due to the much larger molecular size).  Reference dipoles computed with B3LYP/daDZ.}
    \label{fig:alpha_helix_breakdown}
\end{figure}

\begin{figure}[ht]
    \centering
    \includegraphics[width=\textwidth]{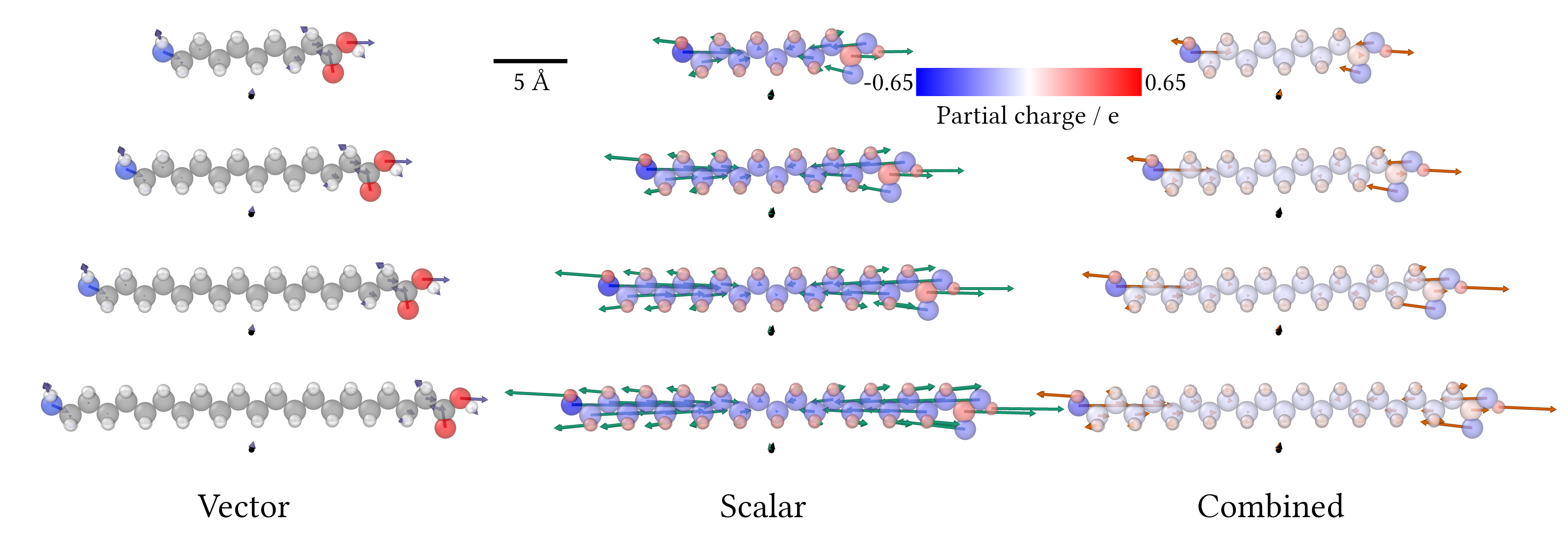}
    \caption[Per-atom breakdown of $n$-amino carboxylic acid predictions]{Per-atom breakdown of the MuML predictions on the $n$-amino carboxylic acid challenge series.  Unlike with the unsaturated version of this set shown in the main text, the total dipole is small and nearly constant with size.  Reference dipoles computed with B3LYP/daDZ.}
    \label{fig:alpha_helix_breakdown}
\end{figure}